\documentclass[usenatbib,usegraphicx,times]{mn2e}
\include{epsf}

\title[Multi-wavelength Cosmological Simulations of Elliptical Galaxies]
{Multi-wavelength Cosmological Simulations of Elliptical Galaxies}
\author[D.\ Kawata and B.K.\ Gibson]
 {D.~Kawata and B.K.~Gibson
\thanks{E-mail: dkawata,bgibson@astro.swin.edu.au}
\\
  Centre for Astrophysics and Supercomputing, 
  Swinburne University of Technology, Hawthorn VIC 3122, Australia
}
\date{Accepted .
      Received ;
      in original form }

\pagerange{\pageref{firstpage}--\pageref{lastpage}}
\pubyear{2003}

\begin{document}

\maketitle

\label{firstpage}

\begin{abstract}

We study the chemodynamical evolution of elliptical galaxies and
their X-ray and optical properties using high-resolution cosmological
simulations. Our Tree N-body/SPH code includes a self-consistent treatment of
radiative cooling, star
formation, supernovae feedback, and chemical enrichment. We present a
series of ${\rm \Lambda}$CDM cosmological simulations which trace the
spatial and temporal evolution of abundances of heavy element in both the
stellar and gas components of galaxies. 
A giant elliptical galaxy formed in the simulations
is quantitatively compared with the observational data 
in both the X-ray and optical regime.
X-ray spectra of the hot gas
are constructed via the use of the {\tt vmekal} plasma model, and analysed
using XSPEC with the XMM EPN response function. Optical properties
are derived by the population synthesis for the stellar component.
We find that radiative cooling is important to interpret the
observed X-ray luminosity, temperature, and metallicity of the 
hot gas of elliptical galaxies.
However, this cooled gas also leads to excessive
star formation at low redshift, and therefore results in underlying galactic
stellar populations which are too blue with respect to observations.
Time variation and radial dependence of X-ray properties and
abundance ratios, such as [O/Fe] and [Si/Fe], of the X-ray emitting
hot gas are also discussed.

\end{abstract}


\section{Introduction}
\label{intro-sec}

 Historically, galactic astronomy was based on optical observation. 
At optical wavelengths, general properties 
especially for nearby elliptical galaxies are well studied.
For example, the various scaling relations for nearby elliptical
galaxies, such as the Colour-Magnitude Relation 
\citep[CMR: e.g.][]{as72,apf81,ble92a,ble92b}
and fundamental plane \citep[e.g.][]{dd87,dlb87},
provide strong constraints on the formation paradigm of elliptical galaxies.
On the other hand, progressive developments of telescopes and instruments 
have opened up other wavelengths, such as X-ray and radio, 
and made it possible to observe galaxies at those wavelengths 
with similar or even higher sensitivity
and resolution than those in the optical observations.
As a result, those multi-wavelength observations 
throw a new light on the study of the formation and evolution 
of galaxies. 

 As for elliptical galaxies, the discovery
of an X-ray emitting hot interstellar medium (ISM) has
dramatically renewed the view of their physical properties and
formation history \citep[e.g.][]{fsjlf79,fjt85}. 
The optical observation represents mainly the properties 
of stellar component. 
On the other hand, X-ray observations offer a unique tool for the study of 
the hot gas component. Since there is little cold gas 
in elliptical galaxies \citep[e.g.][]{bhr92,ae01,ghc01}, 
those two are the major components of elliptical galaxies.
Therefore, any successful scenario of elliptical galaxy formation
must explain {\it both} X-ray and optical observation.

 The aim of this paper is to present our first attempt
to construct a successful self-consistent cosmological 
simulation of elliptical galaxies whose properties are consistent with
both X-ray and optical observations.
To this end, we carry out a series of high-resolution cosmological simulations
using our galactic chemodynamical evolution code \citep[{\tt GCD+}:][]{kg03}.
The code is a three-dimensional tree N-body/smoothed particle
hydrodynamics (SPH) code which incorporates self-gravity, hydrodynamics, 
radiative cooling, star formation, supernovae (SNe) feedback, and metal
enrichment. {\tt GCD+} takes account of
the chemical enrichment by both SNe II and SNe Ia,
and mass-loss from intermediate mass (IM) stars, and follows the 
evolution of the abundances of several chemical elements
in both gas and stellar components. 
Based on the plasma model for hot gas and 
single stellar population (SSP) for stars,
this self-consistent chemodynamical treatment
allows us to derive the X-ray and optical spectral energy distribution (SED)
for the simulation end-products with minimal assumptions.
Taking advantage of this, we compare these results 
for an elliptical galaxy in our simulation with
the observational data at X-ray and optical wavelength
directly and quantitatively, and discuss the physical processes
of the formation and evolution of elliptical galaxies. 

 In the X-ray observation, we focus on the well-studied observational
values of luminosity, $L_X$, temperature, $T_X$, and iron abundance, 
[Fe/H]$_X$. In observed clusters of galaxies, there is well-known
${L_X}-{T_X}$ relation, ${L_X}\propto{T_X}^{b}$:
$b=2.5-3$ \citep[e.g.][]{es91,xw00}. 
On the other hand, the simple spherical top-hat density perturbation theory
predicts that the virial temperature and density of a cluster
with mass $M$ follow $T\propto M^{2/3}(1+z)$
and $\rho\propto(1+z)^3$ at a redshift of $z$ \citep[e.g.][]{tp93}.
At typical cluster temperatures ($T>2$ keV), free-free emission
dominates the cooling process, and the X-ray luminosity
is expected to be $L_X\propto M \rho T^{1/2}$.
If all clusters followed this simple scaling low,
we would expect $L_X\propto T^2$ at a fixed redshift,
which is a significantly shallower slope than that observed.
Recently, the ${L_X}-{T_X}$ relation for groups of galaxies has
also been examined, and is revealed to have much steeper slope,
$b\sim5$, than that for clusters of galaxies
\citep[e.g.][]{hp00,xw00}. Thus, in lower mass systems,
the agreement between the simple theoretical prediction and 
the observations becomes even worse, i.e., the X-ray luminosity of
low-temperature system is unexpectedly small.
\citet{pcn99} calculated the specific entropy of the gas
for observed groups and clusters, and found an ``entropy floor''
of $S\equiv T/n_e^{2/3}\sim100h^{-1/3}$ keV cm$^{-2}$,
where $n_e$ is the electron number density. 
This entropy floor reduces the X-ray
luminosity by lowering the gas density progressively more in smaller
system, and hence reproduces the steep $L_X-T_X$ relations seen in
groups.

 One approach to explain this entropy floor is ``pre-heating''
of the gas before the system collapses \citep{eh91,nk91}.
\citet{lpc00} estimated that a pre-heating temperature of $\sim$0.3 keV
is required to explain the entropy floor, 
if the energy is injected
well before cluster collapse. If energy injection takes place
internally, i.e.\ within collapsed halos, more energy is
required, $\sim 1-3$ keV \citep{ml00,wfn00,bbl01}.
Although the source of the pre-heating is still unclear,
this scenario is also supported by 
numerical simulations which take into account 
the dynamical evolution of the system \citep[e.g.][]{nfw95,mtkp02}.
An alternative mechanism to explain the entropy floor is 
radiative cooling. \citet{gb00} pointed out
that smaller systems, such as groups, have converted more of their 
gas into galaxies than larger systems, and the gas that goes
to form the galaxies was lower in entropy preferentially, thus
raising the mean entropy of the gas that remains.
\citet{vb01} further argued that cooling removes low-entropy gas during
the formation of cluster galaxies, and showed that this mechanism
naturally explains an entropy floor of $\sim$ 100 keV cm$^{-2}$.
Hydrodynamic cosmological simulations also support these analytic
arguments \citep[e.g.][]{ptce00,mtkpc01,mtkp02,dkw02,rv03}.
 
 The size of the potential well of elliptical galaxies corresponds
to the low mass end of the groups observed in the above studies.
\citet[][MOM00]{mom00} 
showed that the X-ray luminosities of elliptical galaxies
are well below that expected from the extrapolation of the
cluster $L_X-T_X$ relation. Therefore, the entropy floor seems to
be still active on the scale of elliptical galaxies.
However, the above theoretical
work focuses on larger systems, such as groups and clusters.
On the elliptical galaxy scale, we must take into account
details of complex radiative cooling processes, star formation,
as well as SNe feedback at the formation epoch and during
the long time evolution of the system \citep{ml86,hhi87}. 
These processes were simplified 
in the above theoretical studies, or were difficult to follow
due to the resolution limit of numerical simulations.
Thus, to examine the $L_X-T_X$ relation for elliptical galaxies 
is a challenging problem, and
this paper will shed light on this issue.

 X-ray observation also provides the chemical composition of
the hot gas, which should contain a fossil record of past
history of the elliptical galaxy evolution, because stellar mass-loss 
products and SNe ejecta are expected to be accumulated in the hot gas.
The standard SNe and stellar mass-loss rates predict
the metallicity of the hot gas to be several times higher than the solar
value \citep[e.g.][]{lm91,amior97,bm99}. 
However, previous measurements of ISM of elliptical galaxies
with {\it ASCA} have shown that the iron abundance is less than 
the solar abundance \citep{amt94,lmt94,mka97,mom00}. This 
discrepancy between the theoretical predictions and observations is called
the ``iron discrepancy'' \citep{amior97}.
X-ray iron abundances remain a controversial
issue. Recently, \citet{bf98,db99,db02} claimed that some X-ray luminous
galaxies have roughly solar abundance, employing a multi-temperature
plasma model. However, even these observed abundances 
are still lower than the theoretical prediction. Also,
high-resolution spectra taken with 
Reflecting Grating Spectrometers (RGS) on board {\it XMM-Newton}
show less than solar abundance \citep{xkp02,spt02,tkmt03}.

To solve this iron discrepancy, a couple of scenarios have been suggested.
Using a chemodynamical evolution model,
\citet{kb98} showed that the metallicity of the ISM 
(especially in the outer regions) becomes lower than that of 
the stellar component in {\it some} elliptical galaxies 
formed by gas-rich disk-disk galaxy mergers, because 
less metal-enriched ISM is tidally stripped away and surrounds
its merger remnant. Considering evolution after the galaxy formed,
using a multi-phase chemical evolution model, 
\citet{ffo96,ffo97} suggested that X-ray emitting hot gas is hardly enriched, 
because the enriched gas cools more easily and drops out of the hot gas phase. 
Unfortunately, there is no study following the chemodynamical evolution
of elliptical galaxies from the epoch of their complex hierarchical formation
to the present, based on the self-consistent cosmological simulation.
Thus, our study would be important to test those scenarios 
in the context of the cosmological evolution.

 From the previous studies above, radiative cooling is 
apparently considered to be a key mechanism in understanding
the X-ray properties, such as $L_X$, $T_X$, and [Fe/H]$_X$,
of elliptical galaxies. Therefore, we first examine the effects
of radiative cooling, comparing the results of simulations with and 
without radiative cooling. Pre-heating is also an interesting mechanism,
although the origin of the pre-heating is unclear. 
Hence, we also consider the effect of SNe feedback which is one of
candidates of the source of the pre-heating as well as the internal
heating within the collapsed halo 
\citep{ml86,hhi87,cdpr91,pcn99,lpc00,wfn00,bbl01}.
In this paper, we would like to clarify the effects of
radiative cooling and SNe feedback on the X-ray properties
of elliptical galaxies, such as $L_X$, $T_X$, and [Fe/H]$_X$.

As mentioned above, we must explain not only the X-ray but also optical 
properties.
Although the previous studies pointed out the importance of
radiative cooling, they did not answer where the cooled gas has gone.
It would be natural to expect that such cooled gas would be the source of
successive star formation. However, the optical observations,
such as the CMR, provide a strong restriction on such recent star formation.
Therefore, we test the CMR of the stellar component of the simulation
end-products as well. 
We will show that this test gives a serious constraint 
on the theoretical model.

The outline of this paper is as follows.
Details of our code, numerical simulation model, and analysis
of X-ray and optical properties
are described in Section \ref{meth-sec}.
In Section \ref{res-sec}, we show the results, and compare
the simulation results with the observational data in both
the X-ray and the optical regime.
Finally, we present a discussion and conclusions 
in Section \ref{dc-sec}.
 Throughout this paper, we adopt for the solar abundance 
the ``meteoric'' values in \citet{ag89}.

\begin{figure*}
\epsfxsize=180mm  
\epsfbox{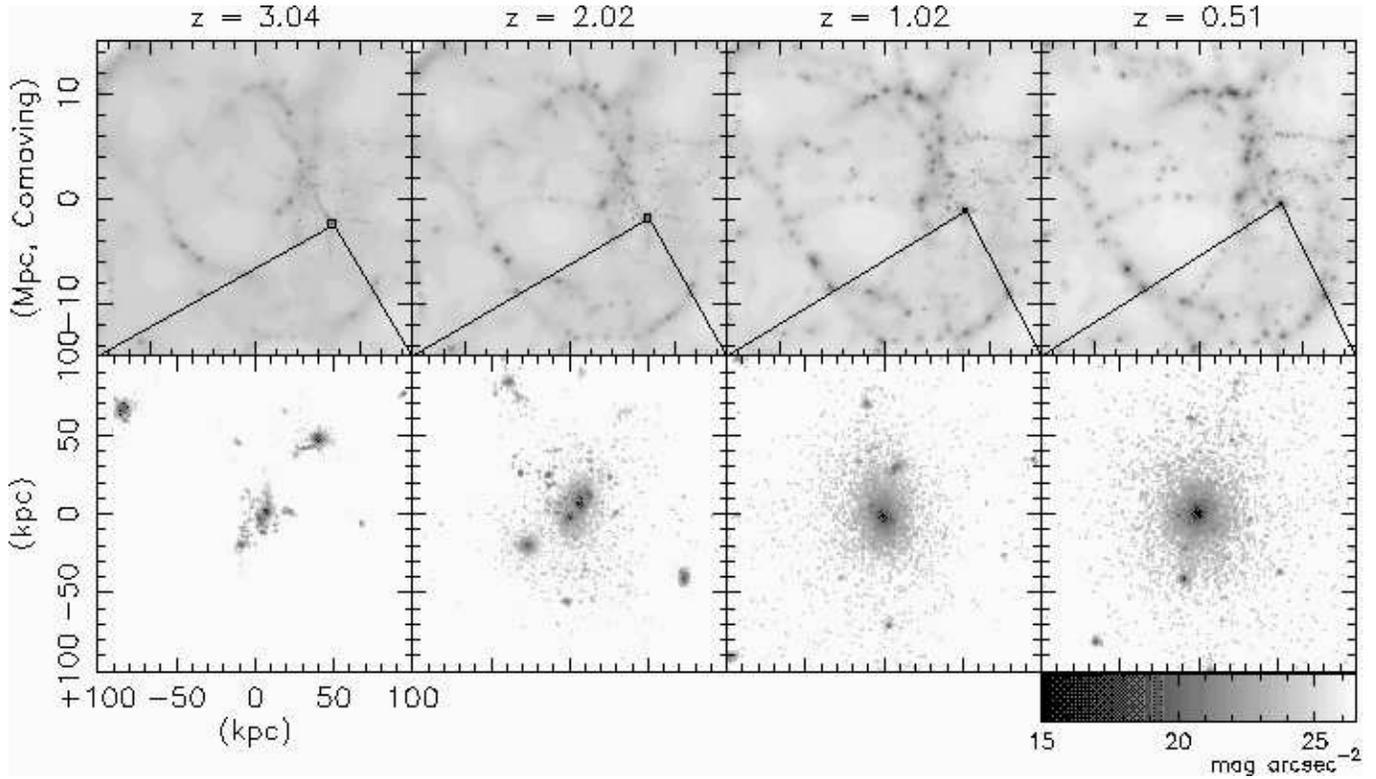}
\caption{Dark matter density map of a portion of the 43~Mpc (comoving) 
simulation volume
({\it upper panels}), and predicted $I$-band image (physical scale)
of the target galaxy
({\it lower panels}), over the redshift range $z$=3.0 to $z$=0.5.
}
\label{evol-fig}
\end{figure*}

\begin{figure*}
\epsfxsize=160mm  
\epsfbox{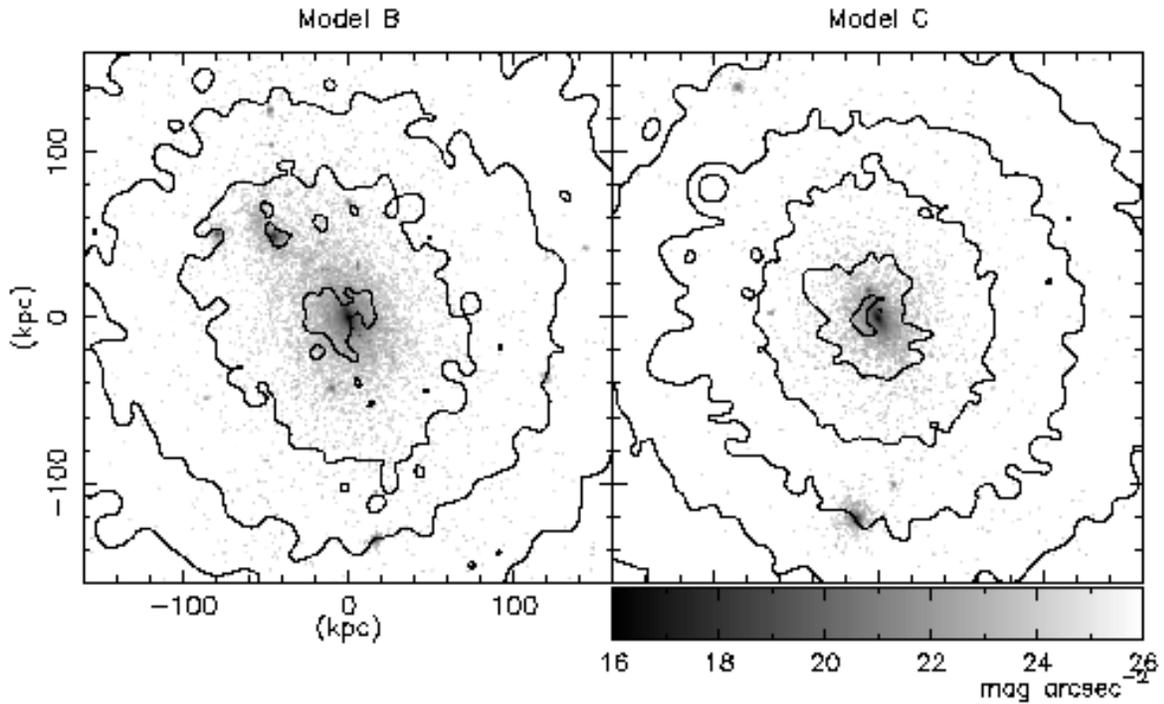}
\caption{ 
 R band image of the target galaxy for Models B (left) and C (right)
at z=0. Overlaid contours show X-ray surface brightness.
}
\label{xopt-fig}
\end{figure*}

\section{Methods}
\label{meth-sec}

\subsection{The Galactic Chemodynamical Evolution Code}
\label{gcd-sec}

 Our simulations were carried out using {\tt GCD+}, our original
galactic chemodynamical evolution code.
Details of the code are presented in \citet{dk99}
and \citet{kg03}.
The code is essentially based on TreeSPH \citep{hk89,kwh96},
which combines the tree algorithm \citep{bh86}
for the computation of the gravitational forces with the SPH
\citep[][]{l77,gm77}
approach to numerical hydrodynamics.
The dynamics of the dark matter (DM)
and stars is calculated by the N-body scheme, and the
gas component is modeled using SPH.
It is fully Lagrangian, three-dimensional, and highly adaptive in space
and time owing to individual smoothing lengths and
individual time steps. Moreover, it includes self-consistently
almost all the important physical processes
in galaxy formation, such as self-gravity, hydrodynamics,
radiative cooling, star formation, SNe feedback, and
metal enrichment. 

Radiative cooling which depends on the metallicity of the gas
\citep[derived with MAPPINGSIII:][]{sd93} is taken into account. 
The cooling rate for a gas with solar metallicity
is larger than that for gas of primordial composition
by more than an order of magnitude.
Thus, the cooling by metals should not be ignored
in numerical simulations of galaxy formation
\citep{kh98,kpj00}.

 We stress that {\tt GCD+} takes account of energy feedback and
metal enrichment by both SNe II and SNe Ia,
and metal enrichment from IM stars 
\citep[see][for details]{kg03}. 
The code calculates the event rates of SNe II and SNe Ia,
and the yields of SNe II, SNe Ia, and IM stars
for each star particle at every time step,
considering the \citet{es55} initial mass function (IMF: mass range of 
0.1$\sim$60 ${\rm M_{\rm \sun}}$) and metallicity dependent stellar lifetimes 
\citep{tk97,ka97}. 
We assume that each massive star ($\geq8\ {\rm M_{\rm \sun}}$)
explodes as a Type II supernova.
The SNe Ia rates are calculated using the model 
proposed by \citet{ktn00}.
The yields of SNe II, SNe Ia, and IM stars
are taken from \citet{ww95}, \citet{ibn99}, and \citet{vdhg97},
respectively.
The mass, energy, and heavy elements ejected from star particles 
are smoothed over the neighbouring
gas particles using the SPH smoothing algorithm.
For example, when the $i$-th star particle ejects the mass of
$M_{\rm SN,{\it i}}$, the increment of the mass of the $j$-th neighbour 
gas particle is given by
\begin{equation}
 \Delta { M_{\rm SN,{\it j}}}
  =  \frac{m_j}{\rho_{{\rm g},i}} { M_{\rm SN,{\it i}}}
  W(r_{ij}/h_{i}),
\label{msneq}
\end{equation}
where
\begin{equation}
 \rho_{{\rm g},i} = \langle \rho_{\rm g}(\mbox{\boldmath $x$}_i) \rangle
 = \sum_{j \neq i} m_j W(r_{ij}/h_{i}),
\end{equation}
and $W(x)$ is an SPH kernel \citep{kg03}.
The simulation follows the evolution of the abundances
of several chemical elements ($^1$H, $^4$He,$^{12}$C, $^{14}$N, $^{16}$O,
$^{20}$Ne, $^{24}$Mg, $^{28}$Si, $^{56}$Fe, and Z,
where Z means the total metallicity).
The self-consistent calculation of both chemical (especially including
the metal enrichment from SNe Ia and IM stars, 
which were often ignored in previous chemodynamical simulations)
and dynamical
evolution by {\tt GCD+} makes it possible for us to analyse
the chemical and physical properties for both the gas and stellar components,
which is essential for this study, 

\subsection{Cosmological Simulation and the Target Galaxy}
\label{cs-sec}

 Using {\tt GCD+}, we have carried out a series of high-resolution
simulations within the adopted standard $\Lambda$-dominated cold 
dark matter ($\Lambda$CDM) cosmology
($\Omega_0$=0.3, $\Lambda_0$=0.7, 
$\Omega_{\rm b}$=0.019$h^{-2}$, $h$=0.7, and $\sigma_8$=0.9).
We use the multi-resolution technique, to achieve high-resolution
in the interesting region, including the tidal forces from the 
large scale structures. 
The initial conditions for the following simulations
are constructed using the public software ``GRAFIC2'' \citep{eb01}.
All our simulations use isolated boundary conditions.
Initially, we perform a low-resolution N-body simulation of a comoving 
30$h^{-1}$ Mpc diameter sphere. Mean separation of the particles is
30$h^{-1}$/48 Mpc. The mass of each particle is $2.90\times 10^{10} 
{\rm M_{\sun}}$, and a fixed physical softening of 18.0 kpc is applied.

 At redshift $z=0$, we selected an 11.6 Mpc diameter spherical region,
which contains a few galaxy-sized DM halos.
We trace the particles which fall into the selected region
back to the initial conditions at $z=38.6$, and identify the volume 
which consists of those particles. Within this arbitrarily shaped volume,
we replace the low resolution particles with 64 times less massive
particles. The initial density and velocities for the less massive 
particles are self-consistently calculated by GRAFIC2, taking into
account the density fields of a lower resolution region.
Finally, we re-run the simulation of the whole volume 
(30$h^{-1}$ Mpc sphere), including the gas dynamics, cooling,
and star formation. The gas component is included only within the
high-resolution region. The surrounding low-resolution region
contributes to the high-resolution region only through gravity.
The mass and softening length of individual gas (DM) particles 
in the high-resolution region are $5.86\times10^7$ 
($3.95\times10^8$) ${\rm M}_{\rm \sun}$ and 2.27 (4.29) kpc, respectively.

 At $z=0$, using the friends-of-friends technique, we found 5 stellar systems
which consist of more than 2000 star particles.
The largest stellar system has an elliptical galaxy like morphology,
and the kinematics shows that the system is supported mainly
by velocity dispersion rather than rotation (the ratio
of the max rotational velocity to the max velocity dispersion, 
$V/\sigma\sim0.7$).
We chose this system as the target galaxy for this study. 
The rest of the paper focuses on this target galaxy. 
The total virial mass of this galaxy is
$M_{\rm vir}\sim2\times10^{13}$~${\rm M_{\sun}}$, similar in size to that of
NGC~4472, a bright elliptical galaxy in the Virgo Cluster.  
The virial mass is defined as the mass within the virial radius,
$r_{\rm vir}$, which is the radius of a sphere containing a mean density of
$178 \Omega_0^{0.45}$ times the critical values, 
$\rho_{\rm crit}=3 {\rm H_0} /8\pi G$, following \citet{enf98}.

As described in Section \ref{intro-sec}, we are interested in
the effect of radiative cooling and SNe feedback.
We carry out a series of simulations
with this initial condition, using the following three models.
Model~A is an adiabatic (i.e.\ no cooling) model;
Model~B includes cooling and weak feedback;
Model~C mimics Model~B, but incorporates stronger feedback.
Model~B assumes that each SN yields the energy of $10^{50}$ ergs,
but Model~C adopts the energy of $10^{52}$ ergs.
The energy in Model~C is unrealistically higher than the canonical
value ($10^{51}$ ergs). We use this model as an extreme comparison case,
to see the dependence of the strength of the SNe feedback more clearly.
Also note that we assume that all kinetic energy of SNe is turned into 
thermal energy on a scale less than our resolution limit,
and thus only the thermal energy of SNe is available to affect 
the surrounding gas of our resolution.

 Fig.~\ref{evol-fig} shows the morphological evolution of 
DM in a central portion of the simulation volume, 
and the evolution of the stellar
component in a 200~kpc region centred on the target galaxy
for Model~B.
The lower panels correspond to the $I$ band (in observed frame)
image of the target galaxy, which is made using population
synthesis techniques as explained in the next section. We take into account 
the K-correction, but do not consider any dust absorption.
The galaxy forms through conventional hierarchical clustering
between redshifts $z$=3 and $z$=1; the morphology has not changed
dramatically since $z$=1.  
Fig.~\ref{xopt-fig} displays the final (at $z=0$) $R$ band image 
(gray scale image) and X-ray surface brightness profile 
(contours) for Models B and C. 
The target galaxy is relatively isolated, with only a few low-mass
satellites remaining at $z$=0. 
Model parameters and final basic properties for each model 
are summarised in Table \ref{mod-tab}.

\begin{table*}
 \centering
 \begin{minipage}{140mm}
 \caption{Model Parameters.}
 \label{mod-tab}
 \begin{tabular}{@{}lcccccccc}
 & energy per SN & $r_{\rm vir} $ &
 \multicolumn{3}{c}{Mass within $r_{\rm vir}$
 (M$_{\rm \sun}$)} & 
 \multicolumn{3}{c}{Number of Particles} \\
Name & (ergs) & (kpc) & Gas & DM & Star & Gas & DM & Star \\
 A\footnote{no cooling or star formation is involved.} & $-$ & 821 &
  $2.14\times10^{12}$ & $1.60\times10^{13}$ & 0 & 36457 & 40492 & 0 \\
 B & $10^{50}$ & 821 &
  $1.45\times10^{12}$ & $1.60\times10^{13}$ & $1.08\times10^{12}$ &
 21136 & 40454 & 18966 \\
 C & $10^{52}$ & 825 &
  $1.83\times10^{12}$ & $1.62\times10^{13}$ & $6.02\times10^{11}$ &
 30765 & 41036 & 10782 \\
 \end{tabular}
 \end{minipage}
\end{table*}

\subsection{Analysis}
\label{ana-sec}

 For all the models, we examine both the resulting X-ray and
optical properties, comparing them quantitatively with observation.
The following sections describe how we derived synthesised spectra
in both X-ray and optical bands from the simulation results. 

\begin{figure}
\epsfxsize=83mm  
\epsfbox{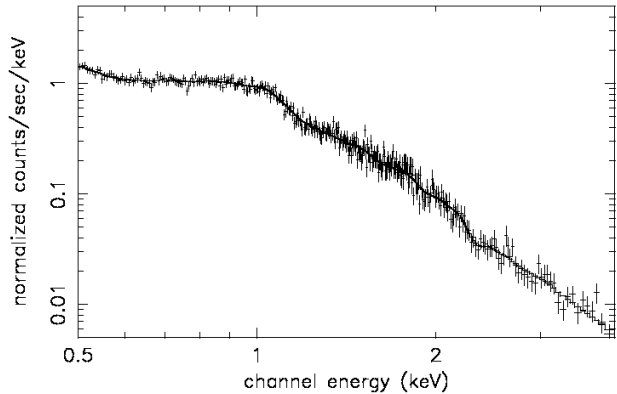}
\caption{ Synthesised X-ray Spectrum (error-bars) 
 and {\tt vmekal} model fit (solid line).
}
\label{xspec-fig}
\end{figure}

\subsubsection{X-ray Synthesised Spectrum and Fitting}
\label{xss-sec}

 To derive the X-ray spectrum, we need to know the 
density, temperature, and abundances of heavy elements
for the gas component. Due to our chemodynamical evolution code,
the gas particles in our simulations carry with them knowledge of 
those physical and chemical properties. 
Using the XSPEC ver.11.1.0 {\tt vmekal} plasma model 
\citep{mgv85,mlv86,jk92,log95}, we derive the X-ray spectra
for each gas particle, and synthesise them within the assumed apertures.
We next generate ``fake'' spectra with the response function of 
the XMM EPN
detector, assuming an exposure time (40~ks) and target galaxy distance
(17~Mpc). 
Finally, our XSPEC fitting of this fake spectrum provides the X-ray weighted
temperatures and abundances of various elements. In this section,
we describe the process of this analysis in detail.

 First, we generate the X-ray spectrum for each gas particle.
We use the subroutine of the XSPEC {\tt vmekal} model.
Input parameters of the {\tt vmekal} model are the hydrogen number density,
temperature, and abundances of He, C, N, O, Ne, Na, Mg, Al, Si,
S, Ar, Ca, Fe, and Ni, with respect to the hydrogen abundance. 
The temperature for each particle are given directly from the 
simulation output. The hydrogen number density is derived 
by the gas density times the hydrogen abundance divided by the
mass of hydrogen atom, i.e.~the proton mass.
Due to the limitation of the size of memory,
our numerical simulation follows the abundance evolution of only
$^1$H, $^4$He,$^{12}$C, $^{14}$N, $^{16}$O, $^{20}$Ne, $^{24}$Mg, 
$^{28}$Si, $^{56}$Fe.
Simply, we assume the abundance of these elements with respect to
$^1$H is the same as the total abundance of all isotopes
with respect to the total ($^1$H and $^2$H) hydrogen abundance, 
e.g.\ $^{16}{\rm O}$/$^1{\rm H}
$=$ {\rm O (=^{16}O+^{17}O+^{18}O)} $/$ {\rm H (=^{1}H+^{2}H)}$.
For the elements whose abundances are not calculated in our simulation, 
we assume [Na/H]$=$[Al/H]$=$[O/H], 
[S/H]$=$[Ar$/$H]$=$[Ca$/$H]$=$[Si/H],
and [Ni/H]$=$[Fe/H], here [X/H]
$=\log_{10}({\rm X}/{\rm H})-\log_{10}({\rm X}_{\sun}/{\rm H}_{\sun})$. 
This assumption should not be an issue, because
we use the same assumption when we fit the synthesised spectrum,
as described later. Finally, we generate
the spectrum for each gas particle within the energy range between
0.1 and 20 keV with 1800 bins, using the {\tt vmekal} model. 

 To compare our results with the X-ray properties observed within a certain
radius, we assume an aperture size. The X-ray fluxes for all the gas particles
within the assumed apetrure radius are added up in each bin,
and the synthesised spectrum is stored as a FITS file to enable XSPEC
to read it as a {\tt model} using {\tt atable} command.
Next, we make ``fake'' spectrum using {\tt fakeit} command in XSPEC.
In this paper, we suppose the observation by EPN detector on {\it XMM-Newton},
and the distance of the target galaxy is 17~Mpc.
We use epn\_ff20\_sY9\_thin.rmf for the EPN response file,
and adopt 40~ks exposure time. Any absorption component or
background are not taken into account.
We binned the spectra to achieve
at least 25 counts per bin, using ``grppha'' tool of HEASOFT ver.5.1.
Fig.~\ref{xspec-fig} shows an
example of fake spectrum with $R=35$~kpc aperture for Model B. 

 Finally, using XSPEC, the generated fake spectrum is fitted
by {\tt vmekal} model within the energy range between 0.5 and 4 keV. 
As mentioned above, here we assume
[Na/H]=[Al/H]=[O/H], [S/H]=[Ar/H]=[Ca/H]=[Si/H],
and [Ni/H]=[Fe/H]. We fix the values of [He/H], [C/H], and [N/H]
to be solar, i.e.~zero, because those abundances do not affect the X-ray
spectrum within the energy range which we use.
Consequently, the free parameters for the fitting
are the temperature and abundances of [O/H], [Ne/H], [Mg/H], [Si/H], 
and [Fe/H]. As shown later, except for Model~A, two temperature
{\tt vmekal} models provide a better fit than one temperature models,
irrespective of the aperture. In a two temperature model, we assume
that the hot and cold components have the same abundances,
to reduce the number of fitting parameters.
The solid line in Fig.~\ref{xspec-fig} shows the best fit 
two temperature {\tt vmekal} model for the fake spectrum of Model~B
with an aperture $R=35$ kpc.
For Model~A, we set the metal abundance to be zero,
i.e.~[X/H]$=-\infty$, because, as there are no stars,
no metals are produced,
and find that a one temperature model can fit the spectrum very well.
Such spectrum fitting gives us the X-ray weighted
temperature, and the abundances of various elements. 

Due to this analysis,
we take into account the line emissions, and the energy range in fitting.
This is more sophisticated way than a simple Bremsstrahlung-based model,
where X-ray luminosity is assumed to be proportional to $\rho^2 T^{1/2}$,
which is conventionally used in numerical simulation studies
of X-ray properties of clusters of galaxies \citep[e.g.][]{nfw95}.
Some previous studies used {\tt mekal} model of XSPEC
 \citep[e.g.][]{me01,trsp02}
or \citet{rs77} model \citep[e.g.][]{kw95,mtkp02}
to derive the X-ray properties. 
Unfortunately, they fixed the metallicity to be the assumed value, 
because of the lack of information about chemical evolution, and
they did not do any fitting of the generated spectrum.
\citet{ffo96,ffo97} used {\tt meka} model of XSPEC to derive 
the X-ray spectrum from the results of their one-dimensional chemical evolution
model, and get the temperature and abundances
by fitting the spectrum using XSPEC. However, they
assumed the solar abundance pattern, i.e.~[Fe/H]=$\log(Z/Z_{\sun})$,
[X/Fe]=0, $Z$ is the total metal (elements heavier than He) abundance,
and $Z_{\sun}$ is the solar abundance.
Thus, the combination of our chemodynamical simulation
and this analysis offers a higher level comparison between 
theoretical models and observational data than that in the previous studies. 
The dependence of the results on the analysis method
is briefly discussed in Section \ref{xgrad-sec}.

 In the following sections, we compare the simulation results
with the X-ray observational data in MOM00. MOM00 measured
the various X-ray properties, such as luminosity, temperature,
metallicity, within an aperture radius of four times the 
$B$-band half-light radius. 
Unfortunately neither the half-light radius nor the optical luminosity
of our models agree with the observed ones, because of a large population
of young stars in the central part of the simulated galaxy
(see Section \ref{opt-sec} for details). 
Since we cannot fix the aperture using the same way as the observation
due to this problem,
we show the results with different apertures, 
and discuss the effects of aperture.
For this purpose, 20~kpc and 35~kpc radius apertures are used.
The former one roughly corresponds to four times the $B$ band
half-light radius for the simulation end-products,
when the light from young ($t_{\rm age}<8$ Gyr) stars are ignored.
The latter one is similar to four times the $B$ band half-light radius
of NGC~4472 ($r_{\rm h}\sim8.5$~kpc). These apertures are applied
to the projected distribution of the gas particles within the virial
radius of the target galaxy at $z=0$.
We use an arbitrary direction for projection, and the same projection
is used when we derive the optical properties.
We have checked that the results we show in this paper
do not depend on the direction of the projection.

 In this paper, we show the X-ray luminosity in
the 0.5--10 keV pass band, because MOM00 showed the X-ray flux in
this pass band. We focus on the X-ray emission
from the hot ISM. The X-ray spectrum derived with
the above analysis is an idealised spectrum from the hot ISM,
because the observed X-ray flux comes from not only the hot ISM but also the 
background and the point sources, such as X-ray binaries.
 MOM00 carefully reduced the background,
and also estimated the flux of the hard component which is considered to 
come from low-mass X-ray binaries. We assume that
the flux of the soft component measured in MOM00 would be dominated by the 
contribution from the hot ISM, and use them in the following section. 
Nevertheless, it is important for this type of study to identify
the X-ray sources, and evaluate the X-ray flux from the hot ISM
correctly using high spatial and spectral resolution observation \citep{sib01}.


\subsubsection{Optical Population Synthesis}
\label{ops-sec}

 The optical photometric properties which are examined in this paper
are derived using the same procedure as that used and
described in \citet{dk01a,dk01b}. Here, we briefly describe this procedure. 
In our simulations, the star particles each carry 
their own age and metallicity ``tag'', 
due to the self-consistent calculation of the chemodynamical
evolution. This enables us to
generate an optical-to-near infrared SED for
the target galaxy, when combined with our population synthesis code
adopting the SSP of the public spectrum and chemical evolution code,
KA97 \citep{tk97,ka97}. 
It is worth noting that our simulation assumes the same shape
and mass range of IMF and the same lifetime of stars as 
those assumed in the SSPs provided by KA97.
Here, the SED of each star particle
is assumed to be that of an SSP, i.e.\
a coeval and chemically homogeneous assembly of stars.
Since the observational data which our results should be compared with
provide the luminosity distribution projected onto a plane,
we have to derive the projected distribution of the SED
from the three-dimensional distribution of star particles.
Finally, we obtain the images projected in the direction used
in the X-ray analysis. 
For example, Fig.~\ref{xopt-fig} shows the $R$-band images
(gray scale image) for Models B and C, 
where a 1000 $\times$ 1000 pixel mesh is chosen 
to span a square region of side 320 kpc, and
the flux of each star particle is smoothed using a gaussian 
filter with the filter scale of 1/4 of the
softening length of the star particle.
These images provide similar information to the imaging data obtained in
actual observations. Thus, we can obtain various photometric properties from
these images in the same way as in the analysis of observational imaging data.
 In the following analysis, we use images similar to
the ones displayed in Fig.~\ref{xopt-fig},
but employing a 750 $\times$ 750 pixel mesh
to span a squared region of side 100 kpc.
The global photometric properties, such as the total luminosity and
the colours, are obtained from the projected image data.
In this paper, we focus on the colour and total magnitude of the target
galaxy. For simplicity, we assume that the luminosity within an
aperture raius of 50 kpc is the total luminosity. The colours are measured
using the same aperture size as the observation which we compare with.
These optical properties are discussed in Section \ref{opt-sec}.

\begin{table*}
 \centering
 \begin{minipage}{140mm}
 \caption{Model Results of X-ray Spectrum Fitting.}
 \label{mres-tab}
 \begin{tabular}{@{}lcccccccc}
 & \multicolumn{8}{c}{Aperture $R<20$ kpc} \\
 &  $L_X$\footnote{X-ray flux (0.5-10.0 keV).} & Fe/H & $kT_1$ & $kT_2$ & 
  EM$_1$/EM$_2$\footnote{Ratio of emission measure of the cooler
  and hotter component.}
 & $\chi^2/\mu$\footnote{degrees of freedom.} &
N$_{\rm g}(R)$\footnote{Number of gas particles within the aperture.} &
N$_{\rm g,hot}(R)$\footnote{Number of hot ($T>10^6$ K)
gas particles within the aperture.}\\
Name & (10$^{41}$ erg s$^{-1}$) & (solar) & (keV) & (keV) & \\
 A\footnote{1T model fit is applied, and abundance is fixed to 0.}
& 30.0 & 0.0 &  $0.45^{+0.00}_{-0.00}$ & 
$-$ & $-$ & 531.9/511 & 1607 & 1606 \\
 B & 0.333 & $0.17^{+0.04}_{-0.03}$ & 
 $1.52^{+0.10}_{-0.05}$ & $0.73^{+0.09}_{-0.07}$ &
 $0.323^{+0.103}_{-0.072}$ & 260.5$/$273 & 245 & 146 \\
 C & 2.87 & $0.21^{+0.01}_{-0.01}$ & 
$1.22^{+0.02}_{-0.01}$ & $0.47^{+0.03}_{-0.03}$ &
 $0.169^{+0.013}_{-0.010}$ & 479.1$/$486 & 438 & 372 \\
 & \multicolumn{8}{c}{Aperture $R<35$ kpc} \\
 A$^f$
& 47.1 & 0.0 &  $0.45^{+0.00}_{-0.00}$ & 
$-$ & $-$ & 478.9/543 & 3657 & 3654 \\
B & 0.710 & $0.12^{+0.02}_{-0.02}$ & 
$1.39^{+0.08}_{-0.07}$ & $0.67^{+0.03}_{-0.04}$ & 
 $0.451^{+0.103}_{-0.093}$ & 311.7$/$347 & 542 & 411 \\
C & 5.48 & $0.12^{+0.01}_{-0.01}$ & 
$1.16^{+0.01}_{-0.01}$ & $0.54^{+0.03}_{-0.02}$ & 
 $0.229^{+0.015}_{-0.011}$ & 549.7$/$572 & 1166 & 1058 \\

 \end{tabular}
 \end{minipage}
\end{table*}

\begin{figure}
\epsfxsize=83mm  
\epsfbox{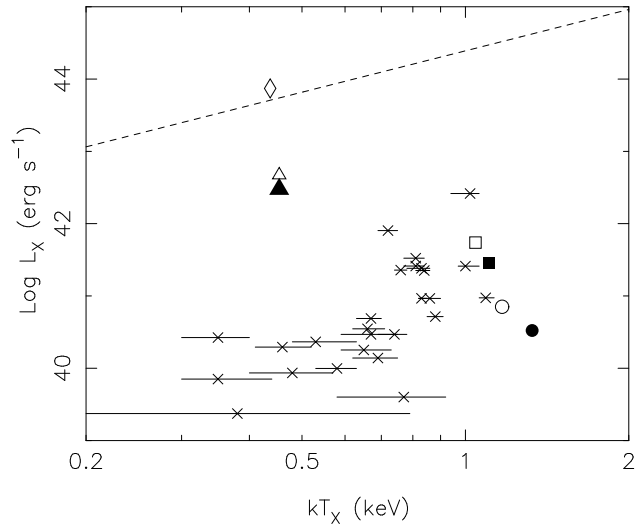}
\caption{ Comparison of the simulated and observed
(crosses with error-bars) ${L_X-T_X}$ relations.
The observational data are taken from MOM00.
The triangle/circle/square indicates the predictions of Model~A/B/C.
The open (filled) symbols denote the values evaluated within the radius
of 35 (20) kpc. 
The dashed line represents the relation obtained from the adiabatic
simulation of \citet{mtkpc01}. 
Open diamond shows the value for Model A
evaluated by the same method as \citet{mtkpc01}.
}
\label{lxt-fig}
\end{figure}

\begin{figure}
\epsfxsize=83mm  
\epsfbox{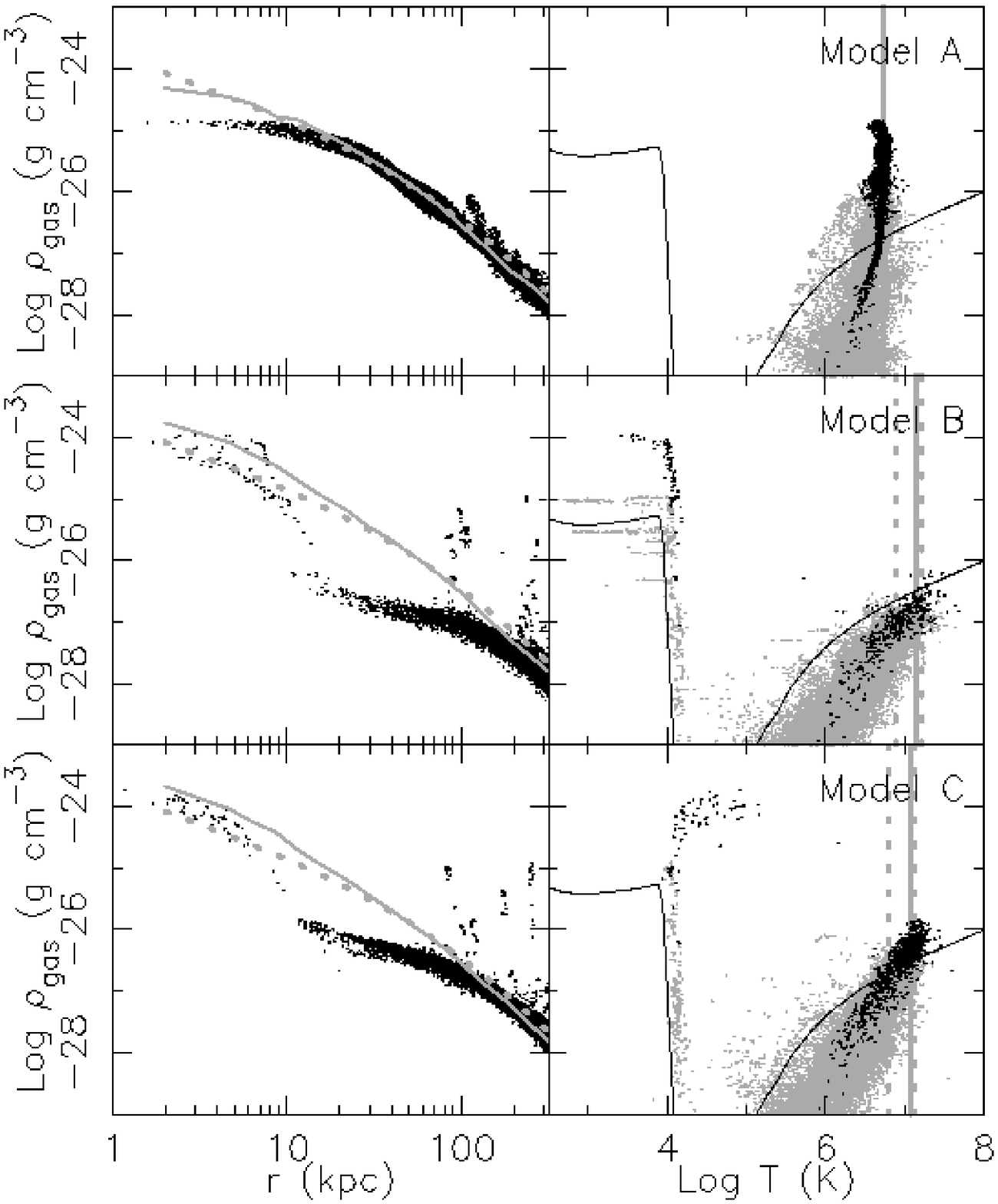}
\caption{ Density vs radius (left) and density vs
temperature (right) distributions of gas particles 
for Model~A (upper), Model~B (middle), and Model~C (bottom).
In the left panels, the gray thick solid lines show 
the DM density profiles, and the gray dotted lines show 
the NFW profiles (see text for details).
The DM densities are scaled by $\Omega_b/(\Omega_0-\Omega_b)$
to compare with the gas density profiles.
The solid curves in the right panels separate the
region where the cooling time (${\rm t_{cool}}$)
is shorter (upper region) and longer than the Hubble time (${\rm t_H}$).
In the right panels, the black (gray) dots represent the particle
within (outside of) the projected radius of 35 kpc.
The vertical gray solid lines show the mean temperatures weighted
by emission measure of the two temperature components (described by 
dotted lines) obtained by the spectrum fitting. Since the spectrum for Model~A 
is fitted by one component, only the solid line is appeared.
}
\label{tcdyn-fig}
\end{figure}

\section{Results} 
\label{res-sec}

 First, we discuss the global properties of the target galaxy
in different models. Section \ref{xrayp-sec} examines
the X-ray properties, and compares them with the observed 
X-ray luminosity$-$temperature (${L_X-T_X}$) relation and
X-ray luminosity$-$iron abundance ($L_X-{\rm [Fe/H]}_X$) relation of elliptical
galaxies in MOM00. Section \ref{opt-sec} discusses the optical
properties of the target galaxy. Here, we focus on the CMR,
which is well-established by the previous observation, and
gives strong constraints on the stellar population of the 
elliptical galaxy. In Section \ref{xgrad-sec}, we mention the X-ray properties
at different radii, i.e.\ the radial gradients. 
Then, the dependence of the derived X-ray properties 
on the analysis method is briefly argued.
We also show time variation of X-ray properties in Section
\ref{evol-sec}. Finally, abundance ratios of the X-ray emitting hot gas 
are discussed in Section \ref{abrat-sec}.

\subsection{X-ray Properties at the Central Region} 
\label{xrayp-sec}

\subsubsection{${L_X-T_X}$ Relation}
\label{lxtr-sec}

 Fig.~\ref{lxt-fig} shows the predicted ${L_X-T_X}$ relation for 
the three models at $z=$0;
crosses with error-bars represent the observational data from MOM00.
As shown in Table \ref{mres-tab} and Section \ref{xss-sec},
except for Model~A, the two temperature {\tt vmekal} model is applied
to fit the synthesised X-ray spectrum. Thus, Fig.~\ref{lxt-fig}
plots the mean temperatures weighted by the emission measure of
the two components for Models B and C.

 The adiabatic model (Model~A) appears incompatible with
the observational data due to its excessive luminosity and low
temperature. The X-ray luminosity is too high, and/or
the X-ray weighted temperature is too low.
The dashed line of Fig.~\ref{lxt-fig} shows the ${L_X-T_X}$
relation obtained from the adiabatic cosmological simulation 
of \citet{mtkpc01}. The result of Model A is rather close to
this line than the observational data. Note that \citet{mtkpc01}
derive the luminosity with a simple analytic formula, and
do not apply any aperture or X-ray pass band, i.e.\
the summation of the cooling rates for all the gas particles with 
temperature larger than 10$^5$ K.
Thus, their luminosity means the bolometric luminosity,
which is the reason why the luminosity for Model~A is smaller than
the dashed line. The way to derive temperature is also different
between \citet{mtkpc01} and us. The temperature 
derived by our analysis should be higher than those by their analysis, 
because of the aperture and X-ray pass band which we apply.
In fact, the open diamond in Fig.~\ref{lxt-fig} shows the result
for Model~A derived by the same way as \citet{mtkpc01}, and
are in good agreement with the dashed line.

 On the other hand, the inclusion of radiative cooling leads to lower 
luminosities and higher temperatures. As a result,
models with cooling (Models B and C) are in better agreement with
the ${L_X-T_X}$ relation of the observed elliptical galaxies.
These results are qualitatively consistent with those of 
\citet{ptce00}, \citet{mtkp02}, and \citet{dkw02} 
\citep[but see also][]{yjs00,lbk00}.

Fig.~\ref{tcdyn-fig} shows the effect of cooling more clearly.
The left panels show the density profiles of gas and DM.
The DM profiles are the spherically averaged density profiles.
We confirm that the DM density profile for Model~A is similar to
the universal profile suggested by \citet[][NFW profile]{nfw97}
for this virialised mass halo and the cosmology which we adopted.
The concentration parameter, $c$, of NFW profile
is a function of the mass of the system, which is represented with
M$_{200}$, the mass within the radius of a sphere
containing a mean density of 200 $\rho_{\rm crit}$.
Using NFW routines provided by 
Julio Navarro\footnote{http://pinot.phys.uvic.ca/$^{\sim}$jfn/charden/},
we obtain the concentration parameters of 
$c=6.72, 6.71, 6.70$ for Models A, B, and C, whose masses are 
M$_{200}=1.62, 1.66, 1.67 \times 10^{13}$ M$_{\rm \sun}$, respectively.
We also ran an N-body only ($\Omega_{\rm b}=0$) simulation 
using the same initial condition,
and confirmed that the DM density profile is identical to that in Model A.
Thus, the inclusion of the adiabatic gas dynamics does not change the
dynamics of the DM. On the other hand, Models B and C have
a steeper DM density profile in the central region ($r<20$ kpc).
This is due to the gravitational drag by the central baryon concentration
which is induced by radiative cooling.
These conclusions are consistent with the previous studies
\citep[e.g.][]{lbk00,ptce00,yjs00}, although those previous studies
focus on larger systems, such as clusters of galaxies whose
mass is $M_{\rm vir}>10^{14} {\rm M_{\sun}}$.

 The gas density profiles are shown with dots 
for all the gas particles within the virial radius.
The gas density profiles are smooth, except for some
particles with high density at large radii which 
are due to gas within the satellite galaxies seen
in Fig.~\ref{xopt-fig}.
In Model~A, the gas density follows the DM density in the outer region. 
However, in the inner
region, the gas density profile becomes flatter than that of the DM,
and the gas densities are smaller than the DM density.
The radius where flattening happens ($\sim20$ kpc)
is much larger than the softening length. Therefore, this is considered
to be caused by the difference in dynamics 
between DM and gas whose thermal pressure
keeps the equilibrium density lower 
\citep[see also][]{nfw95,enf98,ptce00}.

On the other hand, the inclusion of radiative cooling in Model~B
changes the gas density profile dramatically.
The gas density profile no longer follows the DM density profile
within a radius of 200 kpc.
Between this radius and 10 kpc the gas density is much less than
the expected density from the DM density 
($\rho_{\rm DM} \Omega_b/(\Omega_0-\Omega_b)$),
and only recovers that density within 10 kpc.
Consequently the central gas density for Model~B becomes
higher than that for Model A. This is the effect of radiative
cooling, and the right panels of Fig.~\ref{tcdyn-fig} show 
this effect clearly.
The right panels of Fig.~\ref{tcdyn-fig} show the density and
temperature of the gas particles. The lines in the right panels 
show the density
and temperature where the cooling time (${\rm t_{cool}}$)
equals the Hubble time (${\rm t_H}$) at $z=0$, 
i.e.\ the gas above this line can cool within Hubble time.
In the adiabatic case of Model~A, the hydrostatic equilibrium condition
of the gas is determined independently of this line.
Comparing Model~B with Model~A, the inclusion of radiative cooling 
ensures that the temperature of the gas in the upper region of the line
decreases to $T<10^4$ K where the cooling becomes inefficient.
In other words, cooling changes hot gas to
cold gas. The cold gas loses thermal energy and falls further into 
the potential well. This cold dense gas corresponds
to the high density gas within 10 kpc in the left panels 
for Model~B. This conversion of the hot gas to the cold gas
decreases the density of the hot gas in the outer region, and
makes less dense hot gas in the radius between 10 and 200 kpc
in Model~B. Because the cold gas ($T<10^4$~K) does not produce
any emission on the observed X-ray band, this decrease in
the hot gas density is the reason why radiative cooling
drives down the X-ray luminosity.
Moreover, to keep the cooling time long,
the denser gas has to be hotter in Model~B.
This is the reason why radiative cooling drives up the 
X-ray weighted temperature. 

 On the other hand, Model C has similar ${L_X}$ and ${T_X}$
to Model~B.
These are caused by the effect of radiative cooling described
above. However, Model C has slightly lower ${L_X}$ and
higher ${T_X}$ than Model B. The reason for this is also
seen in Fig.~\ref{tcdyn-fig}. Comparing Model~C with Model~B,
some fraction of the hot gas in Model~C stays above the line
of ${\rm t_{cool}=t_{H}}$. 
This is due to the stronger heating by SNe which is
balanced with radiative cooling, and makes the net cooling time for
this gas longer. Therefore, the hot gas in Model~C is 
allowed to be slightly denser and cooler than those in Model~B. 
As a result, ${L_X}$ (${T_X}$)
for Model~C becomes slightly higher (lower) than those for Model~B.

 Fig.~\ref{lxt-fig} shows the X-ray luminosities and temperatures
measured in the different apertures. In Model~A, temperature
is independent of the aperture, which means that the X-ray emitting 
hot gas has little radial temperature gradient within 35 kpc, as seen
in Fig.~\ref{tcdyn-fig}. On the other hand, in Models~B and C,
the smaller aperture results in the higher temperature.
Fig.~\ref{tcdyn-fig} also shows that the hot gas ($T>10^6$ K)
with higher density has higher temperature. In other words, there is
a negative temperature radial gradient in the models including
radiative cooling. We will come back to this issue in Section \ref{xgrad-sec}.

 Finally, we comment on the relationship between the actual temperature
distribution of the hot gas particles and the temperatures
obtained by the spectrum fitting. Right panels of Fig.~\ref{tcdyn-fig} 
show the two temperatures ($kT_1$ for higher temperature;
$kT_2$ for higher temperatuer) as well as the mean value
($kT_m$) obtained by the spectrum fitting within the projected
radius of $R=35$ kpc.
These panels show that $kT_m$ corresponds to
the temperature of the hot gas with high density, and $kT_1$ and $kT_2$
represent the range of the temperature of the hot high-dense gas. 
Although there are a significant amount of the gas 
particles with the temperature less than $kT_2$, their density is
too low to contribute to the X-ray spectrum.
Fig.~\ref{tcdyn-fig} also shows that 
the two components of $kT_1$ and $kT_2$ are induced by the projection
effect.
In fact, we confirmed that a single temperature
model is good enough to fit the X-ray spectrum from the gas particles
within the three-dimintional radius of $r=35$ kpc
\footnote{ Throughout this paper, we use lower-case $r$ for three-dimensional
 radius, and upper-case $R$ for projected two-dimensional radius.}.
This is consistent with the deprojected analysis of the
recent high spatial resolution observations \citep{xkp02,mbfb02,blbm03}.

\begin{figure}
\epsfxsize=83mm  
\epsfbox{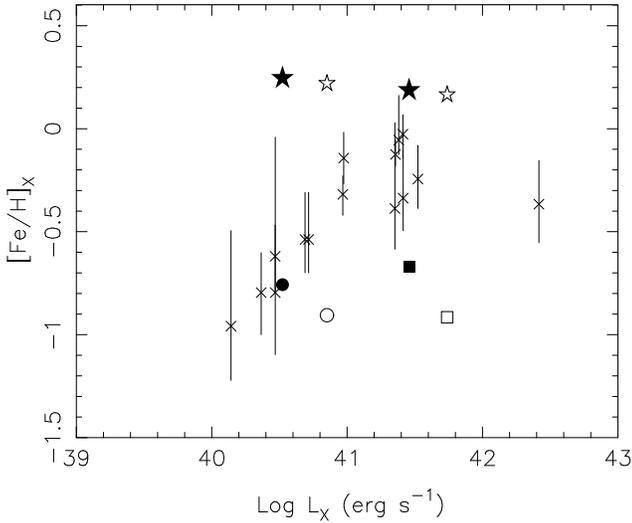}
\caption{ Comparison of the simulated and
observed (crosses with error-bars) ${L_X-{\rm [Fe/H]}_X}$ relations.
The circle (square) shows the result of Model~B (C);
stars show the mean metallicity of the stellar component
for Models~B and C.
The open (filled) symbols denote the values evaluated within the radius
of 35 (20) kpc.
}
\label{lxfe-fig}
\end{figure}

\begin{table*}
 \centering
 \begin{minipage}{140mm}
 \caption{Mass fraction of the ejected iron captured by each component,
 and mass fraction of each component within $r=35$ kpc.}
 \label{mejfe-tab}
 \begin{tabular}{@{}lccccccc}
 & \multicolumn{4}{c}{
  M$_{i, \rm Fe}$\footnote{Iron mass of each component within $r=35$ kpc.}
  $/$M$_{\rm ej,Fe}$\footnote{Iron mass ejected from stars within
   $r=35$ kpc until $z=0$.}}
 & \multicolumn{3}{c}{
  M$_{i}$\footnote{Mass of each component within $r=35$ kpc.}$/$
  M$_{\rm b,tot}$\footnote{Total mass of baryon within $r=35$ kpc.}}\\
Name & star & cold gas & hot gas & escape & star & cold gas & hot gas \\
B &
 0.749 & 0.015 & 0.0008 & 0.235 & 
 0.986 & 0.009 & 0.005 \\
C &
 0.694 & 0.023 & 0.002 & 0.281 &
 0.962 & 0.011 & 0.027 \\
 \end{tabular}
 \end{minipage}
\end{table*}

\subsubsection{$L_X-{\rm [Fe/H]}_X$ Relation}
\label{lxfer-sec}

 Fig.~\ref{lxfe-fig} compares the X-ray weighted
iron abundance of our simulations with the observational data of 
MOM00. As the adiabatic model of Model~A does
not form any stars (not having any cooling), we show
the results only for Models~B and C. 
We measure [Fe/H]$_X=\log {\rm (Fe/H)}_X-\log {\rm (Fe/H)}_{\sun}$ 
using apertures with radii of 20 and 35 kpc.
As explained in Section \ref{xss-sec}, this [Fe/H]$_X$
is derived by X-ray spectrum fitting, and means the X-ray emission
weighted iron abundance of the hot gas. 
The star symbols plotted at the same X-ray luminosities 
as the other symbols show the mean (weighted by mass at z=0) 
stellar metallicity 
within each aperture. Note that we use for the solar abundance the
``meteoric'' values in \citet{ag89}.

 As described in Section \ref{intro-sec}, the iron abundances
observed in the X-ray satellite are significantly lower than
the mean iron abundances of the stellar component.
Since the iron abundances of the hot gas are expected to be higher
than those of stars \citep[e.g.][]{amior97}, this is called
``iron discrepancy'', and one of the biggest mystery of the X-ray
observation of the hot gas of elliptical galaxies.
Our model results show lower [Fe/H]$_X$ of the hot gas, 
compared to their stellar [Fe/H].
Although our model results show even lower [Fe/H]$_X$ than
the observed ones, these are consistent with the observed
iron discrepancy. Therefore, it is interesting to see what has
caused such a low [Fe/H]$_X$ of the hot gas, 
in order to interpret the observed iron discrepancy.

To this end, we examine how much iron is held by each 
component with respect to the amount of iron ejected from the stars.
Table \ref{mejfe-tab} shows those values for
the stars, cold gas, and hot gas, respectively. 
We measure those values within the three-dimensional 
radius of $r=35$ kpc, which encompasses a significant fraction 
of the stellar component.
We found that the deprojected [Fe/H]$_X$ within three-dimensional radius 
of $r=35$ kpc is systematically higher than the projected one.
However, because the difference is small
($\Delta {\rm [Fe/H]}_X\leq0.1$), we conclude that 
the projected iron abundance within $R=35$ kpc is dominated by
the contribution from the central region in three-dimensional space.
Here, the ``hot gas'' is defined as gas with $T>10^6$ K, and
``escape'' means the remainder after subtracting 
the total amount of iron within $r=35$ kpc 
from the total ejected iron from stars within
$r=35$ kpc. Although it is not true that all iron within
this radius comes from the stars within this radius at $z=0$,
it gives us a rough idea where the ejected iron has gone.
Table \ref{mejfe-tab} shows that 
the stellar component holds most of the iron. 
On the other hand, comparing the cold gas with the hot gas,
the cold gas holds more iron than the hot gas, although their masses
are similar. 

\begin{figure}
\epsfxsize=83mm  
\epsfbox{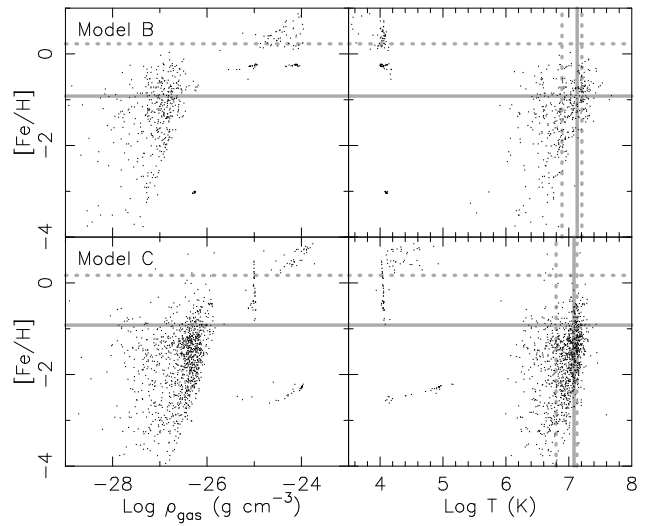}
\caption{ 
 [Fe/H] vs density (left) and [Fe/H] vs
temperature (right) distributions of gas particles 
within the projected radius of $R=$35 kpc
for Model~B (upper) and Model~C (lower).
The horizontal gray solid lines show [Fe/H]$_X$ obtained by the 
spectrum fitting within the radius of $R=35$ kpc.
The horizontal gray dotted lines correspond to 
mean [Fe/H] of stellar component within the radius of $R=35$ kpc. 
In the right panel, the vertical gray solid lines show the mean 
temperatures weighted by emission measure of the two temperature
components (described by dotted lines) obtained by the spectrum fitting
within the radius of $R=35$ kpc.
}
\label{fehrt-fig}
\end{figure}

 This is also demonstrated in Fig.~\ref{fehrt-fig}
which compares [Fe/H]$_X$ obtained by the spectrum fitting with
the distribution of [Fe/H], density, and temperatures for the 
gas particles within the projected radius of $R=35$ kpc.
The value obtained by the fitting, in fact, corresponds to the
[Fe/H] of the hot high-dense gas. The cold gas,
which does not contribute to the X-ray spectrum, has higher
[Fe/H]. Since the stars formed from the cold gas,
the mean [Fe/H] of stellar component is similar to that for
the cold gas.

This metal poor hot gaseous halo is explained by the following mechanism.
Mass-loss from stars and SNe preferentially
enrich the gas in the central region, where the cold gas is 
dominant, because radiative cooling is efficient
(Fig.~\ref{tcdyn-fig}). Also, enriched hot gas can cool more easily
than unenriched gas,
because the cooling is more efficient in the gas with higher metallicity 
\citep{sd93}. Consequently, the cold gas holds more iron
than the hot gas. Moreover, the high density cold gas is incorporated into 
future generations of stars, and thus a large fraction of the iron ejected
from stars is locked into future generation of stars. 
As a result, the hot gaseous halo 
which emits X-ray is not enriched efficiently, 
leading to a lower [Fe/H]$_X$ for Models B and C.
Hence, radiative cooling ensures that the hot gas has a low metallicity.

 We have analysed the mass fraction of the gas of a primordial
origin, i.e.\ the gas which is not ejected from stars,
in the hot ($T>10^6$ K) and cold ($T<10^6$ K) gas within $R=35$ kpc. 
Here, we consider that the mass of the gas of a primoridal origin is
${\rm M_{g,p}}={\rm n_g}\times {\rm m_{g,ini}}$,
where n$_{\rm g}$ is the number of gas particles within $R=35$ kpc, 
and m$_{\rm g,ini}$ is the initial mass of the gas particles, because 
the mass of each gas particle increases only by the mass deposition 
from stars in our modeling of the star formation \citep{dk99}.
As a result, we found that 98 (98) \% of the hot gas
is a primordial origin, while 67 (63) \% of the cold gas is
a primordial origin in Model B (C). This also means that the gas
ejected from stars is locked into the cold gas, and hardly stays
in the hot ISM. In addition, because the hot gas is a primordial origin, i.e.,
the hot gas consists of the gas infalling from the outer region,
the metallicity of the hot gas stays low.

 The aperture effect is also important in [Fe/H]$_X$.
Fig.~\ref{lxfe-fig} shows [Fe/H]$_X$ within the different
apertures, such as 20 and 35 kpc.
In both Models B and C, smaller aperture leads to higher 
[Fe/H]$_{X}$, which means that the hot gas has a negative radial gradient of 
metallicity within 35 kpc. We will discuss
the radial gradients of X-ray properties in Section \ref{xgrad-sec}
in more detail. 

 Finally, we comment on the limitation of our analysis of the
chemical composition for different gas components.
We are not taking into account any thermal conduction or
evaporation process, which is able to change the cold gas to the hot
gas. Since the cold gas is metal rich, if a significant amount of
the cold gas are transfered to the hot gas, the metallicity of the
hot gas would be higher. Unfortunately, there is no good understanding
of how such physical processes are implemented in the numerical
simulation. However, we consider that this process affects in a similar
way to SNe feedback, which is also able to heat the cold gas.
Therefore, our studies comparing different strength of the SNe
feedback should provide us a hint of the total effect of 
the various heating processes.
Also we ignore the mixing of heavy elements among the gas particles. 
If the mixing of heavy elements between the cold and hot gas are
efficient, the metallicity of the hot gas would be higher. 

\begin{figure}
\epsfxsize=83mm  
\epsfbox{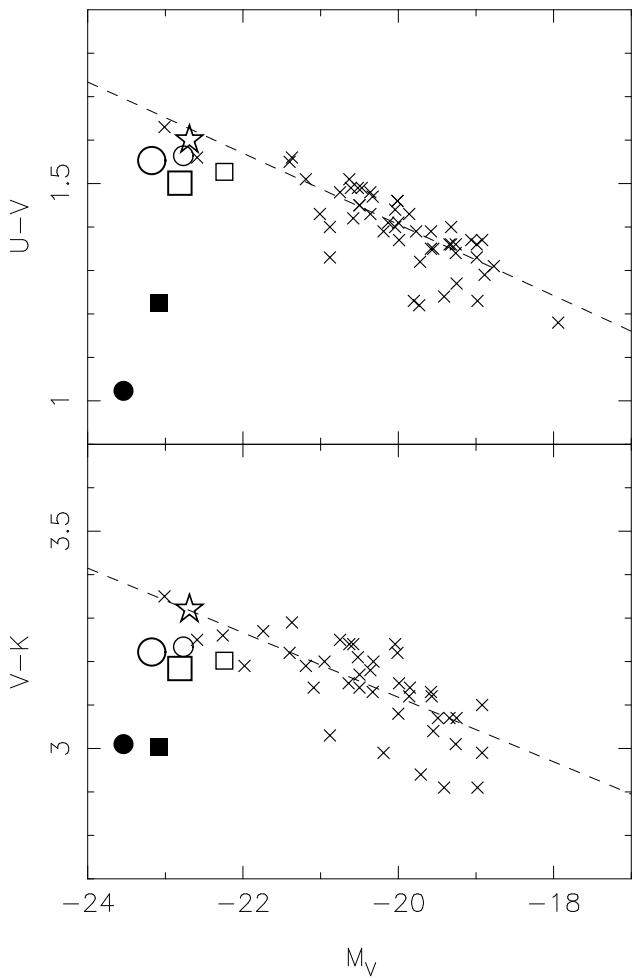}
\caption{ 
 Comparison of the simulated CMRs (filled circle and square for Models~B
and C, respectively) and
that of the Coma cluster ellipticals (crosses).
The dashed line shows the CMR fitted to the Coma Cluster galaxies.
The large (small) open circle and square demonstrate the colour and magnitude
for Models B and C, when the contribution from stars whose age is younger
than 2 (8) Gyr is ignored. Open star shows the position of NGC~4472.
}
\label{cmr-fig}
\end{figure}

\begin{figure}
\epsfxsize=83mm  
\epsfbox{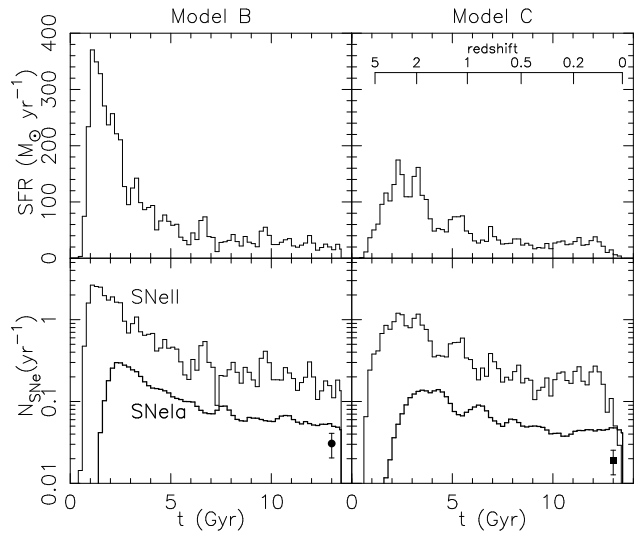}
\caption{ 
Time variation of the star formation rate (upper) and 
the event rate of SNe II (thin lines) and SNe Ia (thick line)
for Models B (left) and C (right).
Solid circle (square) with error-bar 
in lower left (right) panel is taken from the observational SNe Ia 
rate by \citet{cet99}. To show them clearly, those SNe Ia rates
are plotted  at ${\rm t}=13$ Gyr.
}
\label{sfnhr-fig}
\end{figure}

\subsection{Optical Properties} 
\label{opt-sec}

In the last section, we showed that
radiative cooling helps to explain the observed X-ray properties,
such as $L_X$, $T_X$, and [Fe/H]$_X$.  Having said that, any
successful scenario must also explain the optical properties of the
underlying stellar component.
For this purpose, we examine the position of our 
simulated target galaxy in the CMR of observed elliptical galaxies.
The CMR is the simple and well-studied optical scaling relation for 
elliptical galaxies, and gives strong constraints on the formation
history of elliptical galaxies \citep[e.g.][]{ay87,bg97}.
On the other hand, due to our self-consistent treatments of chemodynamics,
the optical properties of the simulation end-products can be
derived with the population synthesis as described in
Section \ref{ops-sec}, and reflect the properties of stellar components,
such as the age and the metallicity. 
Therefore, the CMR provides one of the best tests
for our theoretical model of elliptical galaxies.
Fig.~\ref{cmr-fig} shows the comparison of the simulated galaxies
with the observed Coma cluster galaxies 
as well as NGC~4472 in the $V-K$ and $U-V$ CMR.
The data for the galaxies in the Coma cluster are from \citet{ble92a}.
Since there is no difference between S0s and ellipticals
in the scaling relations which we discuss,
we do not distinguish S0s from ellipticals.
\citet{ble92a} supplies the $U-V$ and $V-K$ colours
which refer to an aperture size of 11 arcsec and
the $V$ band total magnitude derived from a combination of
their data and the literature.
We adopt the distance modulus of the Coma
cluster of $m-M=34.7$ mag: the Virgo cluster distance modulus is 
$m-M=31.01$ \citep{gffkm99} and the relative distance modulus of 
Coma with respect to Virgo is $m-M=3.69$ \citep{ble92b}.
This gives a luminosity distance of 87.1 Mpc for Coma.
We assume that the angular diameter distance equals
the luminosity distance, because the redshift of the Coma
cluster ($z\sim0.023$) is nearly zero cosmologically.
Then the aperture size of 11 arcsec at the distance of the Coma cluster
corresponds to $\sim$ 5 kpc. 
Thus, we also measure the colours within an aperture diameter of 5 kpc
for the simulation end-products.
The total luminosity of the simulation end-products is
defined as the luminosity within the aperture radius of 50 kpc,
which covers almost the whole simulated galaxy (Fig.~\ref{xopt-fig}).

 We can see immediately that the colours of the resulting stellar
components of both Models~B and C are too blue and 
inconsistent with the observational data. This inconsistency can be traced to 
an excessive population of young and intermediate age stars.
The strong feedback in Model C suppresses the formation of young stars, and 
mitigates this problem especially in the $U-V$ CMR,
which is sensitive to a population of young stars.
However, Model C still has too blue a colour to reproduce the
observed colours. Fig.~\ref{sfnhr-fig} shows that the histories
of the star formation and SNe II and Ia rates for all the stars
within $r=35$ kpc.
We can see that significant amounts of stars are continuously 
forming until $z=0$. The SNe II and Ia rates are also too high.
\citet{cet99} provide the observed SNe II and Ia rates of 
nearby elliptical galaxies. Their SNe II rate gives the upper limit
of $0.02\times({\rm H_0}/75)^2$ SNu, and
the rate of SNe Ia which they estimated is 
$(0.018\pm0.06)\times({\rm H_0}/75)^2$ SNu. 
Here, ${\rm H_0}$ is the Hubble constant ( 
${\rm H_0}=70 {\rm km s^{-1} Mpc^{-1}}$ is assumed in this paper)
and SNu is the number of SNe per 100 yr and $10^{10} {L}_{B,\sun}$, 
i.e.~${\rm SNu}=SN (100 {\rm yr})^{-1} (10^{10} {L}_{B,\sun})^{-1}$.
In our models, the SNe II rate is significantly higher than SNe Ia rate,
which obviously contradicts this observation.
In Fig.~\ref{sfnhr-fig} symbols with error-bars show the SNe Ia
rates of \citet{cet99}, which is transfered from units of SNu to
units used in Fig.~\ref{sfnhr-fig} based on the $B$ band luminosity for 
each model. Since the luminosities of both models are too bright,
compared with the observed elliptical galaxies (Fig.~\ref{cmr-fig}),
these rates overestimate the observational data.
Nevertheless, both Models~B and C obviously have too many SNe Ia.

 We also tried to fit the surface brightness profile by 
the Sersic law \citep[see eq.~11 of][]{dk01b}. However, 
we could not obtain any acceptable fit, because the stellar component
is much too centrally concentrated. Such a central concentration of stars is
also induced by radiative cooling which leads to continuous
star formation in the high density central region. Instead, we measured
the half-light radius as a radius where half the total luminosity
is contained, when  the total luminosity is defined as 
the luminosity within the aperture diameter of 100 kpc.
Then, we get $V$ band half-light radii of 4.2 and 3.4 kpc
for Models B and C, respectively. These are too small to reproduce
the observed Kormendy relation, the relation between 
the half-light radius and the luminosity. Therefore, our models
reproduce neither the surface brightness profile nor 
the size of observed elliptical galaxies, 
owing to excessive central radiative cooling.

Therefore, we conclude that although radiative cooling helps to explain the
observed X-ray luminosity, temperature, and metallicity of 
elliptical galaxies, the resulting cooled gas also leads
to the unavoidable overproduction of young and intermediate age stellar
populations, at odds with the observational constraints.
Hence, the test for both X-ray and optical properties
gives stronger constraints on the theoretical model.
We claim that radiative cooling alone cannot explain
both X-ray and optical properties, 
but another physical process to suppress the 
recent star formation induced by cooling is required.

 We found that even our extremely strong feedback assumed in Model C 
cannot stop radiative cooling which leads to the successive star
formation. We compared the X-ray luminosity with the energy 
emitted by SNe II and SNe Ia in Model C, within $r=35$ kpc at z=0.
The X-ray luminosity in the 0.5--10 keV pass band is $3.3\times10^{41}$
erg s$^{-1}$. 
The energy emitted from SNe is estimated using the mean SNe rate
for the last 1 Gyr, and provides $4.4 \times 10^{43}$ erg s$^{-1}$.
The energy of SNe seems larger than the X-ray luminosity.
However, this X-ray luminosity is the luminosity in the 0.5--10 keV
pass band. We measured the X-ray luminosity in the 0.1eV and 1 MeV
pass band, and obtained $2.0\times10^{45}$ erg s$^{-1}$.
This is consistent with the total cooling rate, $1.1\times10^{45}$ 
erg s$^{-1}$, of the gas particles within $r=35$ kpc, which is
calculated with the cooling rate used in our simulation code.
Hence, the total cooling rate is much higher than the SNe
feedback energy. Model C has a higher SNe rate than the observed 
elliptical galaxies (Fig.\ \ref{sfnhr-fig}), and the X-ray luminosity
of Model C is consistent with the observation (Fig.\ \ref{lxt-fig}).
Thus, this result means that in the {\it real} elliptical galaxies
SNe Ia do not generate enough energy to stop cooling.

As expected, if the contribution of the young stars was ignored, 
the resulting colours match the observed CMR. 
Open circles and squares in Fig.~\ref{cmr-fig}
demonstrate that if the star formation in the last 2 Gyr 
did not occur, the CMR of both models would become close to observational
data. If the stars formed in the last 8 Gyr did not exist,
the observed CMR would be roughly recovered by the increase in the colour and
the decrease in the luminosity.
Therefore, one exotic solution is 
to ``hide'' these younger stars within
a bottom-heavy initial mass function (IMF) such that they cannot be
seen today even if they did exist
\citep[e.g.][]{fnc82,mb99}.

Another (more plausible) possibility is that extra heating
sources, such as intermittent active galactic nuclei (AGN) activity, 
suppress star formation
at low redshift.  Before suggesting this is the true solution though,
we must re-examine the predicted X-ray properties of the simulation
end-products after introducing these additional heating sources; we
will be pursuing this comparison in a future paper.

\begin{figure}
\epsfxsize=83mm  
\epsfbox{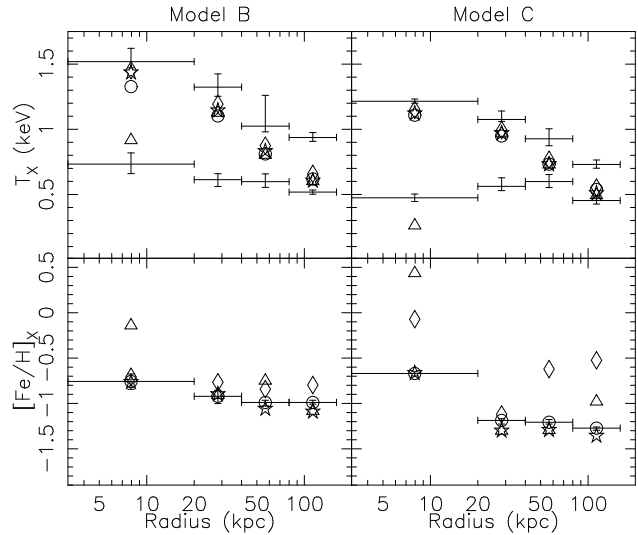}
\caption{ 
 Radial dependences of $T_X$ and [Fe/H]$_X$
for Models B (left panels) and C (right panels).
Different symbols correspond to the values derived by the different
methods, such as spectrum fitting (circles), X-ray luminosity
(0.5$-$4 keV) weighted mean values (diamonds), 
$\rho^2 T^{1/2}$ weighted mean values 
for the gas with $T>12000$ K (triangles), 
and $\rho^2 T^{1/2}$ weighted mean values for the gas with $T>10^6$ K (stars). 
In the upper panels, the error-bars show the two temperatures used in fitting 
the X-ray spectrum,
and circles indicate the mean temperatures weighted by the emission measure of
the two components. The horizontal error-bars show the radius range, and 
the vertical error-bars indicate the 90 \% confidence range obtained
by the spectrum fitting.
}
\label{grad-fig}
\end{figure}

\subsection{Radial Dependences of X-ray Properties}
\label{xgrad-sec}

 In Section \ref{xrayp-sec}, we focused on the mean properties
in the central region. In this section, we discuss X-ray properties
at different radii. First, we briefly mention 
the dependence of the results on the analysis method.
Fig.~\ref{grad-fig} shows the projected radial dependences of $T_X$
and [Fe/H]$_X$ obtained by the following four different methods. Method 1 is 
spectrum fitting (plotted by circles and error-bars ), which 
is used in Section \ref{xrayp-sec}; method 2
takes the mean values weighted by the X-ray luminosity 
predicted by {\tt vmekal} in the $0.5-4$ keV energy range (diamonds); 
method 3 takes the mean values 
weighted by $\rho^2 T^{1/2}$ for the gas with 
$T>12000$ K (triangles); method 4 is the same as method 3 but only 
for the gas with $T>10^6$ K (stars).
We can see that the spectrum fitting gives remarkably similar temperatures
and abundances to method 4.
Therefore, the results derived by the X-ray spectrum fitting
reflect the properties of the high temperatures ($T> \sim10^6$ K) gas.
This is due to the X-ray pass band which is applied in fitting the spectrum
(in this paper, 0.5$-$4 keV, which is a typical range used in the
observational studies). As shown in Fig.~1 of \citet{me01}, 
the shape of the spectrum depends on the temperature of the X-ray
emitting hot gas. As a result, the gas with $T< \sim10^6$ K hardly
contributes to the X-ray spectrum at energies higher than 
0.5 keV. For the same reason, method 3 provides different values from
the ones obtained by method 1.
In particular, the temperature of the central region is well bellow 
the ones obtained by method 1,
due to the overestimate of the contribution from the low temperature
($T< \sim10^6$ K) gas.
Thus, if the simple $\rho^2 T^{1/2}$ weighted mean value is used to compare
the simulation results with the X-ray observational data, 
the temperature limit for the X-ray emitting hot gas is crucial, and 
should be set a temperature as high as $T=10^6$ K.
 
 Except in method 3, the temperature gradient has negative slope.
This is inconsistent with the observed temperature gradients
in elliptical galaxies, which has a positive gradient especially
at small radii \citep[e.g.][]{fj00}. 
The observed positive gradient
in temperature is considered to be evidence for a cooling flow.
On the other hand, since the ${\rm t_{cool}}={\rm t_H}$ line has
positive slope in the density vs.\ temperature diagram
(the right panels of Fig.~\ref{tcdyn-fig}),
including radiative cooling leads to a negative temperature gradient, 
if the gas density profile also has a negative gradient, which 
is naturally expected.
Thus, to explain the observed positive temperature gradient, i.e.\ 
the low temperature ($T<\sim1$ keV) in the central region,
significant amounts of the gas has to stay in the region 
of ${\rm t_{cool}}<{\rm t_H}$ in the density vs.\ temperature diagram 
for a long time, $\sim {\rm t_H}$.
One possible physical process to realise this condition
is again heating sources to balance radiative cooling, although
it requires a fine-tuned physical process.
In fact, the stronger feedback in Model C leads to the shallower
temperature gradient than Model B (see also Fig.~\ref{tcdyn-fig}),
although Model C still fails to reproduce the positive gradient.

 In [Fe/H]$_X$, irrespective of the method, there is a clear negative gradient,
which is consistent with the observational trends of some bright
elliptical galaxies \citep[e.g.][]{fj00}.
This is because stars, which are the source of the iron, 
stay within the central region corresponding to the inner most bin. 
This explains why the smaller aperture gives a higher [Fe/H]$_X$
in Fig.~\ref{lxfe-fig}. 

In [Fe/H]$_X$, method 1 gives similar 
results to method 4.  Particularly in the central bin, method 3 
gives a higher metallicity than method 1, due to the contribution
from the cooler ($T<10^6$ K) gas which is in the central region 
(Fig.~\ref{tcdyn-fig}) and enriched by stars preferentially. 
Surprisingly, method 2 gives higher metallicity than method 1. 
We found that this is because the amount of line emission 
increases with metallicity, and method 2 is biased
towards higher metallicity gas. In fact, if the continuum X-ray 
luminosity is used, and line emissions are ignored,
the luminosity weighted mean values agree with the values 
obtained by the spectrum fitting. Therefore, we should be careful
when we compare the results of theoretical models with the X-ray observations.

\begin{figure*}
\epsfxsize=170mm  
\epsfbox{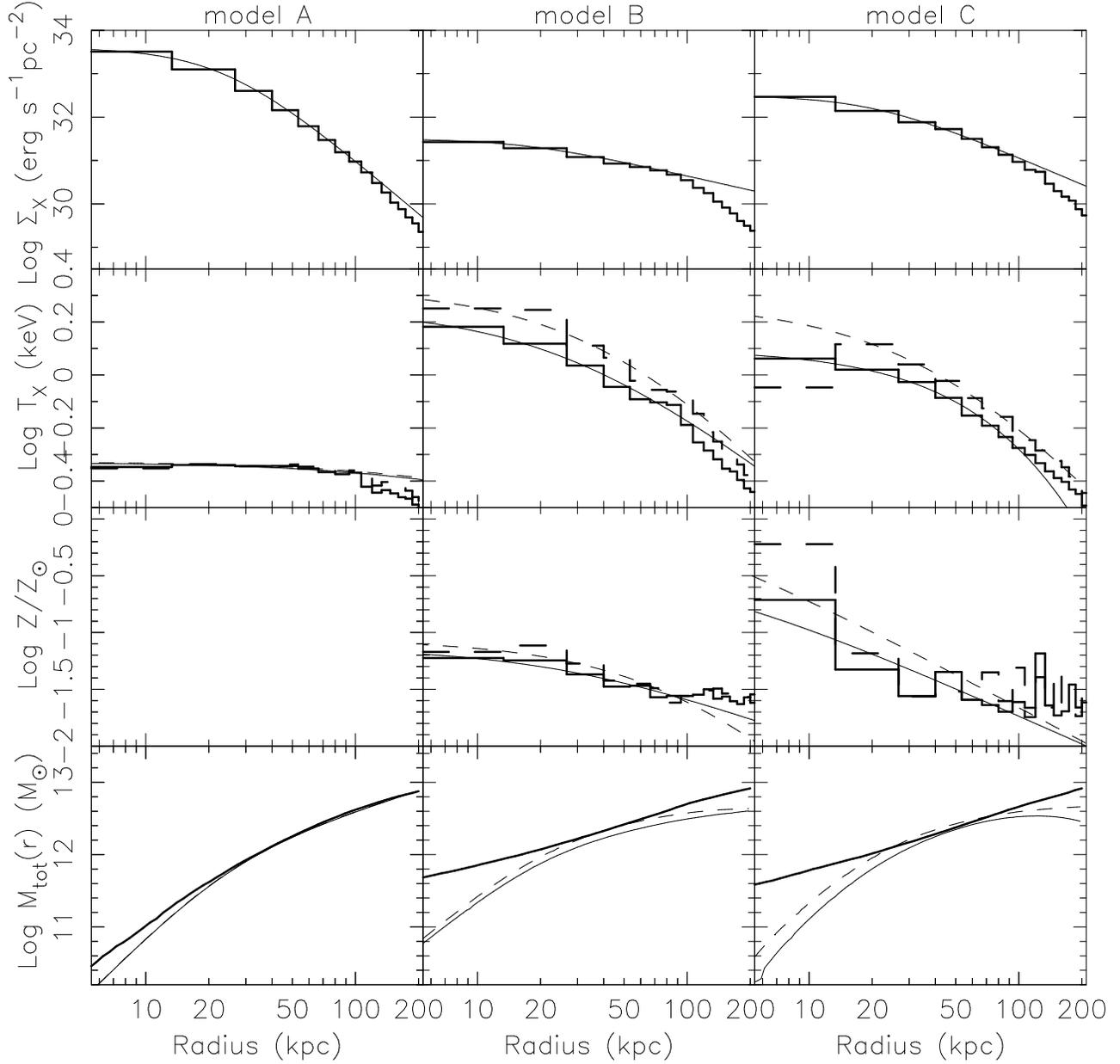}
\caption{ 
Top row: the 0.5--4 keV X-ray radial luminosity profiles (thick lines)
and the best-fit beta profile (thin lines). 2nd row: projected (solid
 lines) and deprojected (dashed lines) temperature
profiles (thick lines) and the best fit profiles (thin lines).
3rd row: the projected (solid lines) and deprojected (dashed lines) 
metallicity profiles (thick lines) and the best fit profiles (thin lines).
Bottom row: ${\rm M_{tot,sum}}(r)$ (thick lines) 
and ${\rm M_{tot,hyd}}(r)$ (thin lines).
The solid and dashed lines corresponds to the profiles calculated 
from projected and deprojected temperature and metallicity profiles,
respectively. 
In model A, the solid and dashed lines are almost overlapped, because
they provides similar values.
}
\label{lxr-fig}
\end{figure*}

 Top panels in Fig.\ \ref{lxr-fig} show the 0.5--4 keV 
X-ray luminosity profile.
We fit these profiles with the follwoing $\beta$ model,
which is used to fit the observed X-ray luminosity profiles 
\citep[e.g.][]{fjt85,km97,mmi98},
\begin{equation}
 \Sigma_X (R) = \Sigma_{X,0} \left[1+(R/R_{\rm core})^2\right]^{-3\beta+0.5}.
\label{betapeq}
\end{equation}
In fitting, we used the data at radii less than 100 kpc,
because the X-ray luminosity drops rapidly at radii further than 100 kpc.
Thin lines in Fig.\ \ref{lxr-fig} show the best fit profile,
and the parameters are summarised in Table \ref{fpara-tab}.
The X-ray luminosity profiles for all the Models are well described
with the $\beta$ model.
The typical values in the observed elliptical galaxies 
are $\beta\sim0.5$ and $R_{\rm core}\sim1$ kpc
\citep[][]{fjt85,km97}. The values of $\beta$ obtained for our models are
similar to the observed values.
It is worth noting that the profile for Model A is
significantly steeper than those of Models B and C.
Thus, radiative cooling leads to a shallower 
X-ray luminosity profiles, i.e., lower $\beta$.
This is also a consequence of the decrease in the central hot gas
density with radiative cooling (Fig.\ \ref{tcdyn-fig}).
On the other hand, the core radius is too large to reproduce 
those for the observed elliptical galaxies, irrespective of models. 
This is also a problem to solve for the current numerical simulation
model. In model A, the core comes from the central region where
the density profile is flat due to the thermal pressure of the adiabatic gas
(Fig.\ \ref{tcdyn-fig}). 
On the other hand, in Models B and C, the large core radii might be caused by 
a poor resolution of the simulation, because in the central region
the number of the hot gas is small due to radiative cooling.

 The X-ray luminosity profile is used to estimate
the total mass, ${\rm M_{tot}} (r)$, within a three-dimensional
radius, $r$, combined with the temperature and metallicity
profiles \citep[e.g.][]{fjt85,mmi98}. Assuming that the X-ray emitting
hot gas is in a hydrostatic equilibrium
and has a spherically symmetric distribution, ${\rm M_{tot,hyd}}(r)$ 
is written by 
\begin{equation}
 {\rm M_{tot,hyd}} (r) = -\frac{k T_{\rm g}(r) r}{G \mu m_p}\left(
\frac{d \ln \rho_{\rm g}(r)}{d \ln r}+\frac{d \ln T_{\rm g}(r)}{d \ln r}
\right),
\label{mreq}
\end{equation}
where $G$, $m_p$, and $k$ are respectively the constant of gravitation, 
the proton mass, and the Boltzman constant; $T_{\rm g}$ and $\rho_{\rm g}$
are the temperature and density of the X-ray emitting hot gas; 
$\mu=0.6$ is the mean molecular weight of the gas. 
Since it is easy to analyse ${\rm M_{tot}}(r)$ from
the output of numerical simulations,
it is interesting to compare the mass profile which is derived using
equation (\ref{mreq}), ${\rm M_{tot,hyd}}(r)$,
with that derived from the simulation outputs directly, i.e.\ the
summation of the mass of particles within the radius, ${\rm M_{tot,sum}}(r)$. 
To use equation (\ref{mreq}), we need the hot gas density and temperature
profile. We derived the hot gas density using the relation
of 
\begin{equation}
\rho_{\rm g}(r)=[\epsilon_X (r)/l_{\rm mekal}(T(r),Z(r))]^{1/2}.
\label{rhogeq}
\end{equation}
Here, $\epsilon_X$ is the X-ray emission per volume,
which is calculated from the following Abel integration 
\citep[e.g.][]{bm98}
of the surface luminosity profile, $\Sigma_X (R)$, of equation (\ref{betapeq}),
\begin{equation}
 \epsilon_X (r) =  -\frac{1}{\pi} \int_{r}^{\infty} 
  \frac{d \Sigma_X(R)}{d R}\frac{d R}{\sqrt{R^2-r^2}}.
\label{abeleq}
\end{equation}
We numerically integrate this equation and obtain $\epsilon_X (r)$.
The expected X-ray emission, $l_{\rm mekal}$ (T,Z), 
is a function of temperature and metallicity, and calculated from
the XSPEC {\tt vmekal} model in the same X-ray pass band as $\Sigma_X$
(0.5--4 keV). Here, we use the metallicity, Z, which is not [Fe/H],
but the total abundance of heavy elements, and assume 
the meteoric solar abundance pattern. Next, the three dimensional
temperature and metallicity profiles are required. In observational
studies, the three dimensional temperature and metallicity profiles are 
often assumed to equal the projected two dimensional profiles. Recently, 
the deprojected three dimensional profiles are also derived, 
assuming a spherical symmetry, and used to estimate 
${\rm M_{tot,hyd}}(r)$ \citep[e.g.][]{mbfb02}.
Therefore, we consider the two cases of the projected profile
and that derived from the three dimensional distribution of particles
(deprojected profile).
The thick lines in the panels of the second and third rows of 
Fig.\ \ref{lxr-fig} show the temperature and metallicity profiles 
of the X-ray emitting hot gas for all the models. 
Here, we use method 4 to derive the profiles, 
because methods 1 and 4 give similar X-ray observed values, as shown above. 
Both temperature and metallicity profiles are fitted with
a power-law function, which is written by 
$T_{\rm g}(R)=T_{\rm g,0} [1+(R/R_T)]^{\alpha_T}$ in the case of the projected
temperature profile \citep[e.g.][]{fjt85}. 
The best fit profiles are presented as thin lines 
in Fig.\ \ref{lxr-fig} and the parameters are summarised in
Table \ref{fpara-tab}. 
In fitting, we used the data at radii less than 100 kpc, which is
the same range as that used in fitting of the luminosity profiles.
Only for the deprojected temperature profiles of Model C,
we exclude the inner most bin whose value is obviously inconsistent with 
the assumed profile. In Model A, the metallicity is set to be zero.
The fitting of the metallicity profiles in Models B and C is 
poor, because the profiles are not smooth. However, the metallicity 
profiles are less sensitive to ${\rm M_{tot,hyd}}(r)$.
Finally, the bottom panels of Fig.\ \ref{lxr-fig} shows
the derived ${\rm M_{tot,hyd}}(r)$ as well as ${\rm M_{tot,sum}}(r)$
which is the summation of the mass of all the components, i.e.\ DM, 
gas, and stars, within the three dimensional radius, $r$, in the simulation
end-products. In Model A, ${\rm M_{tot,hyd}}(r)$ is consistent with
${\rm M_{tot,sum}}(r)$, which is because the X-ray emitting
hot gas in Model A is in a hydrostatic equilibrium. In Models B and C,
around the radius of 50 kpc, ${\rm M_{tot,hyd}}(r)$ matches
with ${\rm M_{tot,sum}}(r)$. 
It means that even if radiative cooling is included,
the status of the X-ray emitting hot gas is close to a hydrostatic equilibrium
in this region. It is not surprising that ${\rm M_{tot,hyd}}(r)$
does not reproduce ${\rm M_{tot,sun}}(r)$ at larger radii ($r>100$ kpc), 
because we exclude those data in fitting. On the other hand, at small radii
${\rm M_{tot,hyd}}(r)$ is systematically smaller than ${\rm M_{tot,sum}}(r)$,
because the hot gas is not in a hydrostatic equilibrium in this region,
due to radiative cooling which continuously converts the hot gas to 
the cold gas. Since our simulations do not reproduce all the
observed properties, we cannot judge whether or not this happens
in the mass estimation from the observational data.
It is also worth noting that ${\rm M_{tot,hyd}}(r)$ from the deprojected
profiles (dashed lines in Fig.\ \ref{lxr-fig}) gives higher
mass and reproduces ${\rm M_{tot,sum}}(r)$ better
than ${\rm M_{tot,hyd}}(r)$ from the projected profiles 
(solid lines in Fig.\ \ref{lxr-fig}) in Models B and C. 
This is because the projected temperature underestimates 
the deprojected temperature,
when there is a significant temperature gradient (Fig.\ \ref{lxr-fig}).
Since the temperature is constant independent of radius, 
in Model A, the projected temperature profile is consistent with 
the deprojected temperature profile, 

\begin{table*}
 \centering
 \begin{minipage}{170mm}
\caption{The results of the fitting of the luminosity, temperature, and
metallicity profiles. The values in round brackets denote the results
of the fitting of the deprojected profiles. }
 \label{fpara-tab}
 \begin{tabular}{@{}lccccccccc}
  & $\log S_{0}$ & $\beta$ & $R_{\rm core}$ 
  & $\log T_{X,0}$ & $\alpha_{T}$ & $R_{\rm T} (r_{\rm T})$ 
  & $\log Z_{0}$ & $\alpha_{Z}$ & $R_{\rm Z} (r_{\rm Z})$ \\
Name & (erg s$^{-1}$ pc$^{-2}$) &  & (kpc) 
 & (keV) &  & (kpc) 
 & ($Z_{\sun}$) &  & (kpc) \\
 A & 33.6 & 0.86 & 23.8 & $-$0.331 ($-$0.328) & 0.15 (0.15) & 128  (141)
   & --    & --   & --   \\
 B & 31.5 & 0.35 & 16.8 &  0.250 (0.327)  & 0.63 (0.88) & 26.8 (47.2)
   & $-$1.14 ($-$1.07) & 0.67 (1.80) & 26.2 (98.7) \\
 C & 32.5 & 0.51 & 20.4 &  0.097 (0.270)  & 3.88 (0.88) & 399  (39.9)
   & $-$0.38 (0.19) & 0.83 (0.98) & 2.36 (1.32) \\
 \end{tabular}
 \end{minipage}
\end{table*}

\begin{figure*}
\epsfxsize=170mm  
\epsfbox{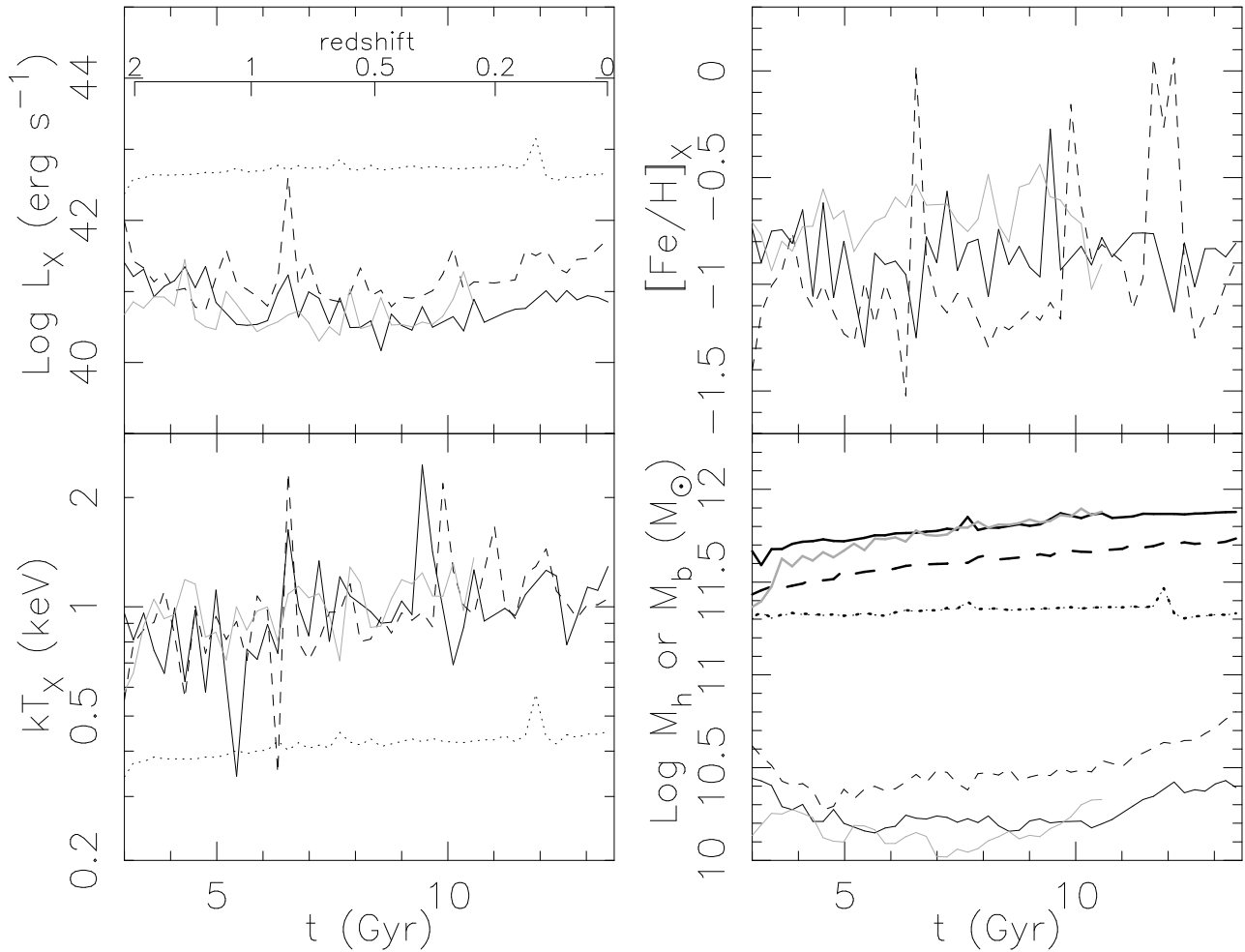}
\caption{ 
Time variation of the X-ray luminosity (upper left), temperature 
(lower left), iron abundance (upper right), and the hot (thin lines)
and baryon (thick line) mass (lower right) within $R=35$ kpc.
The values obtained in Model A/B/C are presented as dotted/solid/dashed
lines. In the lower right panel, since in Model A almost all the baryon
component within this aperture is the hot gas, the thick line
overlap with the thin line. The gray line shows the results
of a higher resolution simulation with the same cooling and feedback
model as Model B.
}
\label{evlx-fig}
\end{figure*}

\subsection{Time Variation of X-ray Properties}
\label{evol-sec}

 In this paper, we focus on the properties of the simulated galaxies
at $z=0$. The advantage of numerical simulation is to be able to
study the evolution of galaxies, while observational data provide
a snapshot of galaxies. Here, we briefly discuss the evolution
history of the simulation end-products.
Fig.\ \ref{evlx-fig} shows the X-ray properties
and the hot gas and baryon mass of the simulated galaxies within $R=35$ kpc
as a function of time. The X-ray luminosities are measured in
the 0.5--10 keV pass band, and temperatures and iron abundances are
measured by method 4 in Section \ref{xgrad-sec}.
Since the number of output files in the simulations are 
limited, the lines are not smooth.
In the X-ray luminosity, all the models have the relatively constant
evolution, except for temporary increases. 
These temporary features are
associated with the accretion of small bound objects,
i.e., minor mergers. In fact, we confirmed that a small bound object
passed through the central region of the target galaxy, and the gas of 
the objects is striped, when the X-ray luminosity
increases temporarily in Model A. It is also seen in Fig.\
\ref{evlx-fig} that at that time the baryon and hot gas 
mass increases, and the temperature also increases.
Although the X-ray luminosities in Models B and C are not smooth,
the X-ray luminosity is always higher in the order of
Models A, C, and B, as seen in Section \ref{lxtr-sec}.
On the other hand, there is significant fluctuation in
the temperatures for Models B and C, and a trend seen in Section 
\ref{lxtr-sec}, i.e.\
Model B has a higher temperature than Model A, represents a condition
in the last 1 Gyr, when the systems are relatively stable.
The mean trend of the iron abundance is constant. However, 
the iron abundance also changes, and gets higher values 
at the same time as the temperature.
We consider that this is due to the SNe feedback
followed by the minor mergers, which heats and enriches the gas.
The baryon mass within this aperture increases continuously
in Models B and C,
because the gas turns into stars and the fresh gas accretes
from the outer region. Since the hot high pressure gas stays 
in Model A, the baryon mass is relatively constant.

 The numerical resolution is always a serious issue for the
galaxy formation simulation, although we have achieved a high resolution
at a similar level to the recent other cosmological simulations which 
include cooling and star formation and follow the evolution until $z=0$
\citep[e.g.][]{mnse03}. Since we focus on the mean properties in the scale
larger than our spatial resolution, we believe
that our conclusion is not heavily affected by the resolution.
To examine this, we carried out a $\sim3.375$ times (by mass) higher
resolution simulation with the same parameters as Model B.
In an initial condition for the higher resolution model, we have assigned
high-resolution particles only in the region surrounding the target
galaxies (roughly within the radius of $4\times r_{\rm vir}$ at z$=0$), 
to reduce the computational time. Therefore, the random seed for
the small density perturbation is different from Model B, which
leads to a different history of minor mergers.
The mass and softening length of each gas (DM) particle
in the high-resolution region are $1.74\times10^7$ ($1.17\times10^8$)
M$_{\rm \sun}$ and 1.51 (2.86) kpc, respectively. 
We stop the simulation at $z\sim0.25$, when
low-resolution particles enter the virial radius of the target
galaxy. The gray lines in Fig.\ \ref{evlx-fig} show the evolution of 
the X-ray properties
and the hot gas and baryon mass of the simulated galaxies within $R=35$ kpc
for the high-resolution model. Although detailed features are different,
the mean values are roughly consistent with those of Model B.
Hence, we conclude that the results presented in this paper are not
sensitive to the numerical resolution.
However, even in this high-resolution model the resolution is still 
much larger than the scales of the various physical processes, such
as the star formation and the SNe feedback. The effect of such
small scale physics on the large scale properties is little known,
and it depends on how to model those processes. 
Therefore, the results shown in this paper might be affected by 
the modeling we adopt. Needless to say, this modeling of
the small scale physics is the most crucial issue for studies
of galaxy formation, which requires further investigations.

\subsection{Abundance Ratios of the X-ray Emitting Hot Gas}
\label{abrat-sec}

 Since we follow the evolution of the abundances of various heavy
elements, we can analyse the abundance ratios of different elements.
Especially, the ratios of $\alpha$ elements and Fe
provide useful information about the chemical enrich  
history, 
because $\alpha$ elements, such as O and Si, are mainly produced
by SNe II, while Fe is mainly produced by SNe Ia.
Table \ref{osife-tab} shows the abundance ratio of O and Si with 
respect to Fe (O/Fe and Si/Fe) for the X-ray emitting hot gas,
the cold gas, and the stellar component within the radius of $R=$35 kpc.
For the hot gas, the values obtained by method 1 and method 4
are shown. For the cold gas and stars, the mean values weighted by the
mass are presented.
We found that Si/H and O/H measured by method 1
are a little different from the values measured by method 4,
which causes differences in O/Fe and Si/Fe between methods 1 and 4.
We consider that this is because
of a difficulty of the X-ray spectrum fitting due to line blending.
Nevertheless, both methods give less than solar O/Fe and
Si/Fe. This is consistent with the recent {\it XMM-Newton} RGS
observations of O/Fe and Si/Fe in the central region of elliptical galaxies
\citep[e.g.\ O/Fe$=$0.4 solar in NGC~4636\footnote{We assumed that
\citet{xkp02} used the photospheric solar abundances, and converted
their results to the values with respect to the meteoric solar abundances.}, 
O/Fe$=$0.37 and Si/Fe$=$1.04 solar in NGC~5044;][]{xkp02,tkmt03}.
Table \ref{mzej-tab} shows the mass fraction of Fe, O, and Si
ejected from SNe II, SNe Ia, and the IM stars within the last 1 Gyr
for Models B and C at $R<35$ kpc.
Table \ref{mzej-tab} also presents O/Fe and Si/Fe of the total ejected gas.
It appears that SNe Ia are a dominant source of Fe,
which induces the low O/Fe and Si/Fe. Higher
Si/Fe than O/Fe is also explained by a significant contribution
of SNe Ia which produce a fair amount of Si, compared to O. 
However, note that the rates of SNe II and SNe Ia are too high
in both Models B and C, as discussed in Section \ref{opt-sec} 
(Fig.~\ref{sfnhr-fig}). 
Especially, the SNe II rates in the observed elliptical galaxies are 
very small, and thus our high SNe II rate overestimates O and Si production.
In addition, Table \ref{osife-tab}
shows that O/Fe and Si/Fe for the stellar component are solar or less, 
while the observed bright elliptical galaxies have over solar abundance
ratios ($\alpha$/Fe$\sim$2). 
This is also because stars continuously form from the cold gas which has
low O/Fe and Si/Fe (Table \ref{osife-tab}).
This low $\alpha$/Fe for the stellar component 
underestimates O/Fe and Si/Fe for the mass-loss from the IM stars,
although the contribution of the mass-loss of IM stars 
to the metal enrichment is small, compared with SNe.
Hence, strictly speaking, the chemical enrichment history of our
simulation is inconsistent with the observation. 
Nonetheless, broadly speaking, our simulation demonstrates that
the observed low Si/Fe and O/Fe in the hot ISM of elliptical galaxies
are explained by the dominant contribution of SNe Ia to the chemical 
enrichment of the X-ray emitting hot gas.


\begin{table*}
 \centering
 \begin{minipage}{140mm}
 \caption{Oxygen and Silicate abundances with respect to iron for
X-ray emitting hot gas, cold gas, and stellar component at $R<$35 kpc.}
 \label{osife-tab}
 \begin{tabular}{@{}lcccccccc}
 & \multicolumn{4}{c}{hot gas} &  \multicolumn{2}{c}{cold gas} &
 \multicolumn{2}{c}{stars}\\
 & \multicolumn{2}{c}{method 1} & \multicolumn{2}{c}{method 4} & 
 \multicolumn{2}{c}{(T$<10^6$ K)}\\
 & O/Fe & Si/Fe & O/Fe & Si/Fe & O/Fe & Si/Fe & O/Fe & Si/Fe \\
Name & (solar) & (solar) & (solar) & (solar) & (solar) & (solar)
 & (solar) & (solar) \\
B & 0.21 & 0.22 & 0.20 & 0.51 & 0.63 & 0.91 & 1.02 & 1.13 \\
C & 0.91 & 0.64 & 0.60 & 0.82 & 0.51 & 0.80 & 0.89 & 1.07 \\
 \end{tabular}
\end{minipage}
\end{table*}

\begin{table*}
 \centering
 \begin{minipage}{140mm}
 \caption{Ejected heavy element mass fraction from SNe II, SNe Ia,
and IM stars and the abundance ratios of the total ejected gas.
The values denote a mean values within the last 1 Gyr from $z=0$
at $R<35$ kpc.}
 \label{mzej-tab}
 \begin{tabular}{@{}lccccccccccc}
 & \multicolumn{3}{c}{SNe II} &  \multicolumn{3}{c}{SNe Ia} &
 \multicolumn{3}{c}{IM stars} &
 \multicolumn{2}{c}{total} \\
 & f$_{\rm ej,Fe}$\footnote{f$_{\rm ej,Fe}={\rm M_{ej,Fe,SNeII}/M_{ej,Fe,tot}}$}
 & f$_{\rm ej,O}$ & f$_{\rm ej,Si}$
 & f$_{\rm ej,Fe}$ & f$_{\rm ej,O}$ & f$_{\rm ej,Si}$
 & f$_{\rm ej,Fe}$ & f$_{\rm ej,O}$ & f$_{\rm ej,Si}$
 & O/Fe & Si/Fe \\
B & 0.29 & 0.74 & 0.52 & 0.50 & 0.02 & 0.25 & 0.21 & 0.24 & 0.23
  & 0.62 & 0.90 \\
C & 0.19 & 0.62 & 0.40 & 0.63 & 0.04 & 0.36 & 0.19 & 0.34 & 0.24
  & 0.48 & 0.77 \\
 \end{tabular}
\end{minipage}
\end{table*}

\section{Discussion and Conclusions}
\label{dc-sec}

 Our cosmological chemodynamical simulations make it possible to undertake
quantitative comparisons between the theoretical models
and the observational data in both the X-ray and optical regime
with minimal assumptions. This is the first attempt to explain
both the X-ray and optical properties of observed elliptical
galaxies by self-consistent cosmological simulations.
First, we found that radiative cooling is required to explain the
observed X-ray luminosity, temperature, and metallicity of 
elliptical galaxies. Comparison between models ignoring and
including radiative cooling clarifies that radiative cooling ensures that
the hot dense gas turns into the cold (i.e.~non X-ray emitting) gas,
and keeps X-ray emitting gas at high temperature and low density
(Fig.~\ref{tcdyn-fig}).
As a result, including radiative cooling leads to lower X-ray
luminosity and higher X-ray weighted temperature, and provides
a better agreement with the observed ${L_X-T_X}$ relation 
(Fig.~\ref{lxt-fig}).
Although this effect of cooling has already been shown by previous
studies \citep[e.g.][]{ptce00,mtkp02,dkw02}, our study newly confirms
that radiative cooling works in the same way on galactic scales
using simulations with higher physical resolution.
In addition, we find that stronger SNe feedback leads to higher
X-ray luminosity and lower temperature, because heating by the SNe feedback
makes the cooling time of the hot gas longer, and thus
the gas with higher density and lower temperature
is allowed to stay as X-ray emitting hot gas (Fig.~\ref{tcdyn-fig}). 
Radiative cooling also ensures that the hot gaseous halo has not been
enriched efficiently. Stars preferentially enrich the gas in the central
region, where cooling is efficient. The enriched gas can then cool
easily and be incorporated into future generation of stars.
In fact, we found that a large fraction of iron ejected from stars is
locked into stars. This effect of cooling
has been already suggested by \citet{ffo96,ffo97},
although they did not consider any star formation induced by the cooled gas. 
Our more self-consistent numerical simulation confirms their scenario.

To summarise the X-ray study, our radiative cooling models succeed in 
reproducing the X-ray luminosity, temperature, and metallicity of observed
elliptical galaxies. However, we also found that the resulting cooled gas
leads to unavoidable overproduction of young and intermediate age stellar
populations, at odds with the optical observational constraints.
We examined the position of our simulated galaxy in the observed Coma
cluster CMR, and found that the colours of the resulting stellar
components are inconsistent with the observational data (being too blue,
Fig.~\ref{cmr-fig}).
Therefore, this study demonstrates that the cross check over both 
X-ray and optical properties 
gives stronger constraints on the theoretical models, and is essential
for any successful scenario of elliptical galaxy formation and evolution.
For example, our results throw doubt on the recent claim
that {\it only} radiative cooling is required to explain 
steeper slope of the ${L_X-T_X}$ relations of groups and elliptical
galaxies.

 Consequently, our model fails to reproduce both X-ray and optical 
properties of the observed elliptical galaxies. Nevertheless,
our study provides some hints to explain those observed properties.
The biggest problem of the present model is too many 
recent (z$<$1) star formation (Fig.~\ref{cmr-fig}).
Therefore, one exotic (if somewhat {\it ad hoc}) solution is 
to ``hide'' these younger stars within
a bottom-heavy initial mass function (IMF) such that they cannot be
seen today even if they did exist \citep[e.g.][]{fnc82,mb99}.
A more plausible possibility is additional physical process
which suppresses the cooling and star formation after the system has formed.
Although we assume the extremely strong SNe feedback in Model~C,
it is not enough to suppress the star formation \footnote{
Note that this might come from our poor modeling of the SNe feedback.
The energy feedback from SNe is one of the most difficult
processes to model in galaxy formation simulations. 
Various SNe feedback models have been suggested 
\citep[e.g.][]{nw93,ykkk97,hp99,tc00,sh03}. Unfortunately 
they neglect SNe Ia. 
}.
Hence, the SNe Ia feedback is unlikely to be a candidate of this physical
process, although it should contribute in some degree.
We expect that extra heating by intermittent AGN activity,
which is not included in our numerical model, 
is a major mechanism to suppress the star formation at low
redshift \citep{bm02,bk02}. Recent observations suggest that all 
elliptical galaxies have a central black hole. Therefore,
once the cold gas falls into the central part of galaxy,
the AGN is expected to be active and heat up the
surrounding gas, and might be able to blow out the gas 
from the system via a radio jet.
Once the gas accretion is stopped by AGN heating, the AGN itself
loses fuel and goes back to a quiescent state, which allows the gas
to cool again. 
This self-regulation
cycle might be able to suppress the star formation significantly
for a long time.
In fact, recent high resolution X-ray images taken by {\it Chandra}
reveal that the central hot gas in elliptical galaxies
is not smoothly distributed, but is cavitated on scales comparable to the
radio emission \citep[e.g.][]{bvfen93,fj01,jfv02}.
The X-ray holes are also seen on the galaxy clusters scale ($<50$ kpc),
and are often coincident with extended radio lobes \citep[e.g.][]{mwn00,
atn01,bsmw01,hcrb02}. Those features are considered to be
a relic of the significant influences of AGN activity on the 
hot gas of elliptical galaxies and intra cluster medium \citep{cbkbf01}.

This AGN heating mechanism might also help to solve the problem in
the observed [Mg/Fe]--magnitude relation of elliptical galaxies
\citep[e.g.][]{fm94,mpg98,kg03}.
To explain the [Mg/Fe]--magnitude relation,
larger galaxies are required to stop star formation at an earlier epoch.
Since the recent observations claim that the black hole mass
increases with the velocity dispersions of their host bulges
\citep[e.g.][and reference therein]{mf01},
the heating effect by AGN is expected to increase
with the stellar mass of elliptical galaxies, and 
star formation in the larger galaxies are suppressed more strongly
once the AGN acts.
We are planing to construct self-consistent models of
this mechanism, and examine how such a self-regulation mechanism 
induced by AGN affects the X-ray and optical properties for
elliptical galaxies with different luminosities.

\section*{Acknowledgements}

We thank Frazer Pearce for his helpful advice during
the completion of this manuscript, and the referee,
Michael Loewenstein, for his thorough reading of the manuscript and
constructive comments.
We are grateful to Nobuo Arimoto and Tadayuki Kodama
for kindly providing the tables of their SSPs data.
We acknowledge the Yukawa Institute Computer Facility,
the Astronomical Data Analysis Center of the National Astronomical
Observatory, Japan, the Australian and Victorian Partnership for Advanced
Computing, where the numerical computations for this paper were performed.
This work is supported in part by
the Australian Research Council through the Large Research Grant 
Program (A0010517) and Japan Society for the Promotion of Science
through the Grants-in-Aid for Scientific Research (No.\ 14540221).

\label{lastpage}

\end{document}